\documentclass{aa}
\usepackage{color}

\usepackage[colorlinks,citecolor=blue,linkcolor=blue,urlcolor=blue]{hyperref}

\title{Tension of toroidal magnetic field in reconnection plasmoids and~relativistic jets}
\titlerunning{Toroidal fields in plasmoids and jets}
\author{
Krzysztof~Nalewajko\thanks{\email{knalew@camk.edu.pl}}
}
\authorrunning{Nalewajko}
\institute{
Nicolaus Copernicus Astronomical Center, Polish Academy of Sciences, Bartycka 18, 00-716 Warsaw, Poland
\label{inst_ncac}
}

\abstract
{
Toroidal magnetic field is a key ingredient of relativistic jets launched by certain accreting astrophysical black holes, and of plasmoids emerging from the tearing instability during magnetic reconnection, a candidate dissipation mechanism in jets.
Tension of toroidal field is an anisotropic force that can compress local energy and momentum densities.
We investigate this effect in plasmoids produced during relativistic reconnection initiated from a Harris layer by means of kinetic particle-in-cell (PIC) numerical simulations, varying the system size (including 3D cases), magnetisation, or guide field.
We find that:
(1) plasmoid cores are dominated by plasma energy density for guide fields up to $B_z \sim B_0$;
(2) relaxed `monster' plasmoids compress plasma energy density only modestly (by factor $\sim 3$ above the initial level for drifting particle population);
(3) energy density compressions by factors $\gtrsim 10$ are achieved during plasmoid mergers, especially with the emergence of secondary plasmoids.
This kinetic-scale effect can be combined with a global focusing of the jet Poynting flux along the quasi-cylindrical {bunched spine} (a proposed jet layer adjacent to the cylindrical core) due to poloidal line bunching (a prolonged effect of tension of the jet toroidal field) to enhance the luminosity of rapid radiation flares from blazars.
The case of M87 as a misaligned blazar is discussed.
}


\begin{document}

\maketitle

\section{Introduction}

Extremely luminous, highly variable, non-thermal high-energy cosmic sources of radiation are often interpreted as originating from powerful collimated outflows known as relativistic jets \citep{2016ARA&A..54..725M,2019ARA&A..57..467B}.
In many supermassive active galactic nuclei (AGNs), narrow jets and their relativistic motions can be observed directly \citep[e.g.,][]{2001ApJS..134..181J,2016AJ....152...12L}.
In the subclass of blazars, small viewing angle allows for huge ($\sim 10^5$) relativistic boost of non-thermal radiation produced by particles accelerated during dissipative processes in the approaching jet \citep{1995PASP..107..803U}.
The most luminous blazars, belonging to the subclass of flat spectrum radio quasars (FSRQ), produce $\sim {\rm GeV}$ $\gamma$-ray flares with apparent luminosities $L_{\rm fl,FSRQ} \sim 10^{50}\,{\rm erg\,s^{-1}}$
and variability time scales of $t_{\rm fl,FSRQ} \sim 10\,{\rm min}$ \citep[e.g.,][]{2011ApJ...733L..26A,2016ApJ...824L..20A,2020NatCo..11.4176S}\footnote{However, even for the brightest blazar flares detected in $\gamma$-rays by the {Fermi satellite}, the statistical significance of variability on suborbital timescales (typically within $\sim 30\,{\rm min}$ windows) is limited to $\sim 2\sigma$ \citep{2019ApJ...877...39M}; see also \cite{2017Galax...5..100N}.}.
Another subclass of blazars, the high-energy peaked BL Lac objects (HBL), produce $\sim {\rm TeV}$ $\gamma$-ray flares with apparent luminosities of $L_{\rm fl,HBL} \sim 10^{47}\,{\rm erg\,s^{-1}}$ and variability time scales of $t_{\rm fl,HBL} \sim 3\,{\rm min}$ \citep{2007ApJ...664L..71A}.
Even after accounting for the relativistic boost, the energetic requirements for the blazar zones can be very demanding in terms of local energy density and radiative efficiency (e.g., \citealt{2012MNRAS.425.2519N,2016ApJ...824L..20A}; Appendix~\ref{app_blazar_flares}).

Toroidal magnetic field is an essential ingredient of relativistic jets launched by magnetised accretion onto spinning black holes \citep[e.g.,][]{1984RvMP...56..255B,2020ARA&A..58..407D}.
It is the main carrier of electromagnetic momentum (Poynting) flux (the initial form of jet power) \citep{1976MNRAS.176..465B}.
Its pressure acting along the velocity streamlines is the primary agent of acceleration to relativistic speeds \citep[e.g.,][]{1989ASSL..156..129C,2007MNRAS.380...51K}.
In jet regions that expand laterally, the toroidal field decreases slower than the poloidal field, likely becoming the dominant component at distances of $\sim 0.1 - 1\;{\rm pc}$ typically inferred for emission of blazar flares \citep{2014ApJ...789..161N}.

Toroidal magnetic field is also an essential ingredient of plasmoids (flux ropes) produced during large-scale magnetic reconnection due to the tearing instability of thin current sheets \citep{1963PhFl....6..459F,2007PhPl...14j0703L}.
Relativistic reconnection is one of the primary candidate mechanisms of energy dissipation considered for relativistic jets to explain the non-thermal particle acceleration and broad-band radiation in blazars \citep{2015MNRAS.450..183S}.
In particular, rapid gamma-ray flares of blazars have been modeled in terms of reconnection-driven relativistic Alfvenic outflows -- minijets \citep{2009MNRAS.395L..29G,2011MNRAS.413..333N}\footnote{However, it has been argued that relativistic turbulence provides more consistent beaming statistics than minijets \citep{2012MNRAS.420..604N,2023ApJ...946L..51S}.}, slower and denser plasmoids \citep{2013MNRAS.431..355G,2016MNRAS.462.3325P,2019MNRAS.482...65C}, head-on plasmoid mergers \citep{2015ApJ...815..101N,2024MNRAS.531.4781Z}, or tail-on plasmoid mergers \citep{2020MNRAS.497.1365O}.
Although produced at vastly smaller scales than relativistic jets, reconnection plasmoids can be described by essentially the same physical principles.

A fundamental property of the magnetic field is that the associated stress tensor has an anisotropic component called magnetic tension \citep[e.g.,][]{2007cemf.book.....P}.
For any loop of toroidal field, tension induces a force acting towards the symmetry axis with the effect of compressing the loop interior.
In the absence of other forces, an axisymmetric force-free equilibrium radial profile of toroidal magnetic field is $B_\phi(r) \propto r^{-1}$, implying a strongly concentrated energy density profile of $u_{\rm B}(r) = B_\phi^2(r)/8\pi \propto r^{-2}$.
If such a profile could be maintained between widely separated scales $r_1 \ll r_2$, energy density at the inner radius would be greatly enhanced $u_{\rm B,1}/u_{\rm B,2} \sim (r_2/r_1)^2$.

In relativistic jets, the separation between macrophysical scale (jet width) and microphysical scale (particle gyroradius) can in principle reach $\sim 10^{10}$ (see Appendix \ref{app_scale_separ}).
The potential for energy density enhancement due to magnetic tension is thus enormous.
On the other hand, scale separations that can be realised in numerical simulations are $\sim 10^4$.
Moreover, in the presence of other forces acting away from the symmetry axis, e.g., pressure gradients of plasma or axial magnetic field \citep[e.g.][]{2022ApJ...931..137O}, compression of energy density is reduced and divided between magnetic field and plasma in various proportions.

\vskip 1em

In Section \ref{sec_plasmoids}, we consider the case of plasmoids emerging from reconnecting current layers in relativistically magnetised plasma.
The internal structure of plasmoids can be investigated by kinetic numerical simulations, in particular using the particle-in-cell (PIC) method \citep[e.g.,][]{2016MNRAS.462...48S, 2018MNRAS.481.5687P, 2020MNRAS.497.1365O, 2021ApJ...912...48H, 2023MNRAS.523.3812S, 2023ApJ...943L..29H, 2023ApJ...959..122C}.
Previous numerical studies of reconnection plasmoids (summarised in Section \ref{sec_plasmoids_prev}) adopted various configurations of PIC simulations, e.g., 2D vs. 3D, in periodic or open boundaries, various background magnetisations, relativistically cold or hot plasma, without or with guide field, without or with radiative cooling.

The main result of this work is the analysis of energy density compression in plasmoids produced in a series of PIC simulations of relativistic reconnection, presented in Section \ref{sec_plasmoids_results}.
As described in Section \ref{sec_plasmoids_sims}, these simulations were initiated from hot Harris-type current layers in periodic boundaries, leading to the formation of relaxed `monster' plasmoids.
We checked the effects of guide magnetic field and the third dimension.
While we do not include radiative cooling in those simulations, in Section \ref{sec_plasmoids_compare} we compare our results with previous simulations presented in \cite{2020MNRAS.497.1365O}, which included synchrotron radiative cooling and open boundaries.

\vskip 1em

In Section \ref{sec_jets}, we consider the lateral structure of relativistic jets and the role of toroidal magnetic field.
The context for this is the possibility of producing luminous and rapid flares observed in blazars powered by plasmoids produced by relativistic reconnection.
The key question is whether reconnection can produce plasmoids in the right region across the jet, where the compression of plasmoid cores can be multiplied by the compression of the jet?

Previous studies summarised in Section \ref{sec_jets} indicate that tension of toroidal field competes with other forces (e.g., centrifugal); while it does not dominate jet collimation, it contributes to a gradual differentiation of the jet structure into a quasi-cylindrical inner core and a paraboloidal outer layer.
One of the key effects of toroidal field tension is the poloidal field bunching \citep{2009ApJ...699.1789T}, which effectively focuses the inner loops of toroidal field (and part of the Poynting flux) around the cylindrical jet core, in an adjacent quasi-cylindrical layer that we call the {bunched spine}.
The focusing of toroidal field is expected to destabilize the jet core to current-driven $m=1$ kink modes, seeding reconnecting current layers and plasmoids around the jet core, where compression of plasmoid cores and high-$\sigma$ particle acceleration can be combined most effectively with compression of the inner jet and beamed relativistically to produce rapid and bright gamma-ray flares of blazars.

\vskip 1em

Section \ref{sec_concl} presents the {discussion and} conclusions.
We propose a specific scenario for the production of luminous radiation flares from relativistic jets by combining energy density enhancements due to tension of toroidal magnetic field at the scales of reconnection plasmoids (investigated by means of kinetic PIC simulations) and the global lateral jet structure.
{High-resolution observations of a central ridge along the nearby misaligned jet of M87 may constrain the proposed jet structure.}

\section{Reconnection plasmoids}
\label{sec_plasmoids}

\subsection{Previous studies}
\label{sec_plasmoids_prev}

Physical properties of plasmoids produced during relativistic magnetic reconnection have been studied by means of kinetic (particle-in-cell; PIC) numerical simulations.
Here we report a selection of previous studies initiated from Harris current layers.
These studies adopted various assumptions: 2D or 3D, periodic or open boundaries, cold or hot background plasma, without or with guide field, without or with synchrotron cooling.

The guide field in the problem of magnetic reconnection is the field component that is not reversed in the initial configuration.
As the anti-parallel field component becomes the toroidal field of individual plasmoids, the guide field becomes the axial field concentrated in the plasmoid cores.
In the absence of initial guide field, the plasmoid cores are supported by pressure of plasma, resulting in the Z-pinch configuration.

\vskip 1em

\cite{2016MNRAS.462...48S} measured plasmoid profiles (functions of radius $r$) in 2D simulations ($2L$ domain with $L \sim 3600 d_{\rm e}$ with $d_{\rm e} = c/\omega_{\rm p}$ the skin depth, $\omega_{\rm p} = (4\pi e^2n_0/m)^{1/2}$ the plasma frequency\footnote{Subscripts $0$ refer to the background plasma.}; using open boundaries) with relativistically cold background plasma ($\Theta_0 = k_{\rm B}T_0/mc^2 = 10^{-4}$) without radiative cooling: plasma density was found to scale like $n \propto r^{-1}$, magnetic energy fraction $\epsilon_{\rm B} = u_{\rm B}/(n_0mc^2) \propto r^{-1.2}$ (with $n_0$ the background plasma density), kinetic energy fraction $\epsilon_{\rm kin} = (\left<\gamma\right>-1)n/n_0 \propto r^{-1.4}$; the largest plasmoids of half-width $w \sim L/20 \sim 180 d_{\rm e}$ reached core densities of $n_{\rm c}/n_0 \sim 300 (\sigma_0/10)^{1/2}$, magnetic energy fractions of $\epsilon_{\rm B,c} \sim 800 (\sigma_0/10)^{3/2}$, and kinetic energy fractions of $\epsilon_{\rm kin,c} \sim 4000 (\sigma_0/10)^{3/2}$.

\cite{2018MNRAS.481.5687P} analysed comparable simulations ($2L$ domain with $L \sim 1000 r_{\rm L}$ with nominal `hot' gyroradius $r_{\rm L} = \sigma_0^{1/2} d_{\rm e} \sim 3d_{\rm e}$ for magnetisation $\sigma_0 = 10$; and periodic boundaries) and obtained even larger plasmoids with half-widths of $w \sim 300 d_{\rm e}$, and reported for them magnetic energy fractions of $\epsilon_{\rm B} \sim 20$.
In this and subsequent works of that group, the population of initial drifting particles was excluded from the analysis.

\cite{2020MNRAS.497.1365O} investigated evolution of plasmoids in 2D simulations ($L \sim 1500\rho_0$ domain with open boundaries) in relativistically hot plasma ($\Theta_0 = k_{\rm B}T_0/mc^2 \sim 10^6$, hence background gyroradius $\rho_0 = \lambda_{\rm D}/\sigma_0^{1/2}$ is related to the Debye length $\lambda_{\rm D} = \Theta_0^{1/2} d_{\rm e}$ with magnetisation $\sigma_0 = B_0^2/(4\pi\Theta_0n_0mc^2) \sim 10$), and with synchrotron cooling.
In case of slow cooling ($\Theta_0 = 2\times 10^5$), they obtained plasmoids growing to half-widths of $w \sim 100\rho_0 \sim 30\lambda_{\rm D}$ with core densities of $n_{\rm c}/n_0 \sim 60$ and magnetic field strengths of $B_{\rm c}/B_0 \sim 4$.
In case of fast cooling ($\Theta_0 = 1.25\times 10^6$), the plasmoids were smaller ($w \sim 60\rho_0 \sim 20\lambda_{\rm D}$) with significantly denser cores ($n_{\rm c}/n_0 \sim 300$) and slightly stronger magnetic fields ($B_{\rm c}/B_0 \sim 6$).

\cite{2021ApJ...912...48H} investigated the structure of plasmoids in 2D PIC simulations with background hot magnetisation $\sigma_0$ up to $100$, reaching $w \sim 30 r_{\rm L} \sim 300 d_{\rm e}$ (with $r_{\rm L} = \sigma_0^{1/2} d_{\rm e}$ the hot gyroradius; $d_{\rm e}$ the skin depth). Considering the force balance between magnetic tension and plasma pressure, they characterised the radial profiles of plasmoids with power laws $\rho \propto r^{-1}$ and $B \propto r^{-2/3}$, noting also strong stratification of mean particle energy $\left<\gamma\right>$ (resulting with even steeper profile of plasma energy density).
For $\sigma_0 = 100$ they achieved enhancement of plasma density $\rho$ by factor $\sim 30$ and magnetic field strength $B$ by factor $\sim 3$.

A comprehensive investigation of compression of plasmoid cores in 3D PIC simulations of relativistic reconnection was reported by \cite{2023MNRAS.523.3812S}.
They considered background plasma with hot magnetisation $\sigma_0 \sim 25$ and mildly relativistic temperature ($\Theta_0 = 4$), taking into account synchrotron cooling (regulated by magnetic field strength $B_0$) and guide magnetic field $B_{\rm g}$.
Compression has been analysed in a parameter space of plasma density $n/n_0$ and magnetic field strength $B/B_0$, by which they showed that enhancement of $n$ is not always correlated with the enhancement of $B$, and this has complex implications for the luminosity of associated radiation signals.
In their reference case (3D, radiative with $B_0 = 2\times 10^{11}\;{\rm G}$, moderate guide field $B_{\rm g}/B_0 = 0.4$), the limit to compression of both parameters was identified at $n/n_0 \sim 30$ (comparable with initial density of the drifting plasma $n_{\rm d}/n_0$) and $B/B_0 \sim 2$.
That limit was not significantly different in the non-radiative case ($B_0 = 2\times 10^8\;{\rm G}$), it depended somewhat (but not monotonically) on the guide field $B_{\rm g}$ and the aspect ratio $L_z/L_x$ of the 3D domain.
The density compression increased with increasing magnetisation $\sigma_0$ in proportion to increasing $n_{\rm d}/n_0$.

Effects of strong synchrotron cooling on reconnection plasmoids have been recently investigated by \cite{2023ApJ...943L..29H} in 2D and \cite{2023ApJ...959..122C} in 3D.
Strong synchrotron cooling reduces plasma pressure, causing further plasmoid contraction and plasma compression, it also severely limits the maximum energy that can be reached by the particles. Nevertheless, plasmoid cores are generally the brightest sources of radiation within sites of relativistic reconnection.

\begin{figure*}
\includegraphics[width=\textwidth]{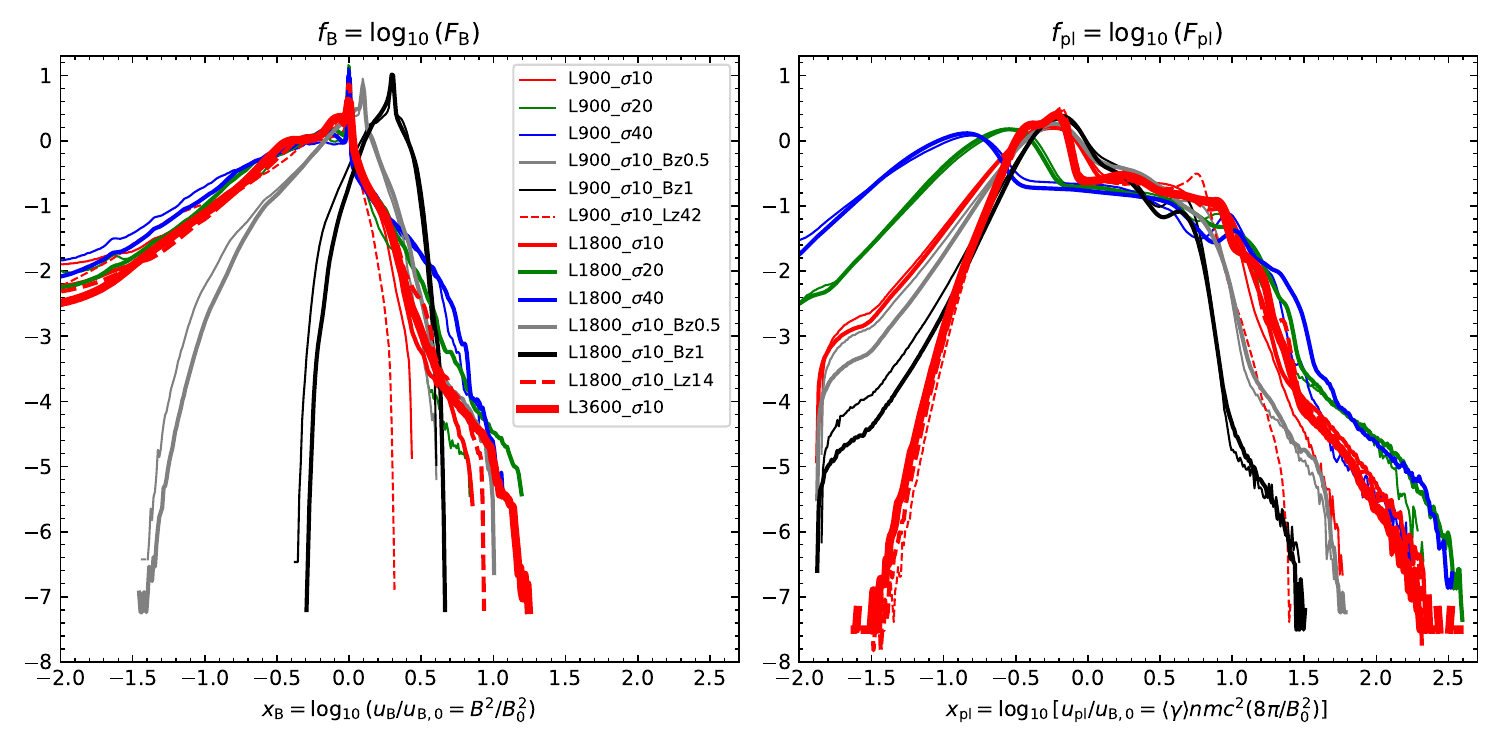}
\caption{Logarithms $f = \log_{10}(F)$ of volume distributions $F(\mu) = {\rm d}F/{\rm d}\mu$ over argument $\mu = \log_{10}(u/u_{\rm B,0})$ with $u$ the energy density: of magnetic fields $u_{\rm B} = B^2/8\pi$ (left panel), and of the plasma $u_{\rm pl} = \left<\gamma\right> n m c^2$ (right panel).
Functions $f(\mu)$ were averaged over the duration of each simulation.}
\label{fig_voldist}
\end{figure*}

\begin{figure*}
\includegraphics[width=0.33\textwidth]{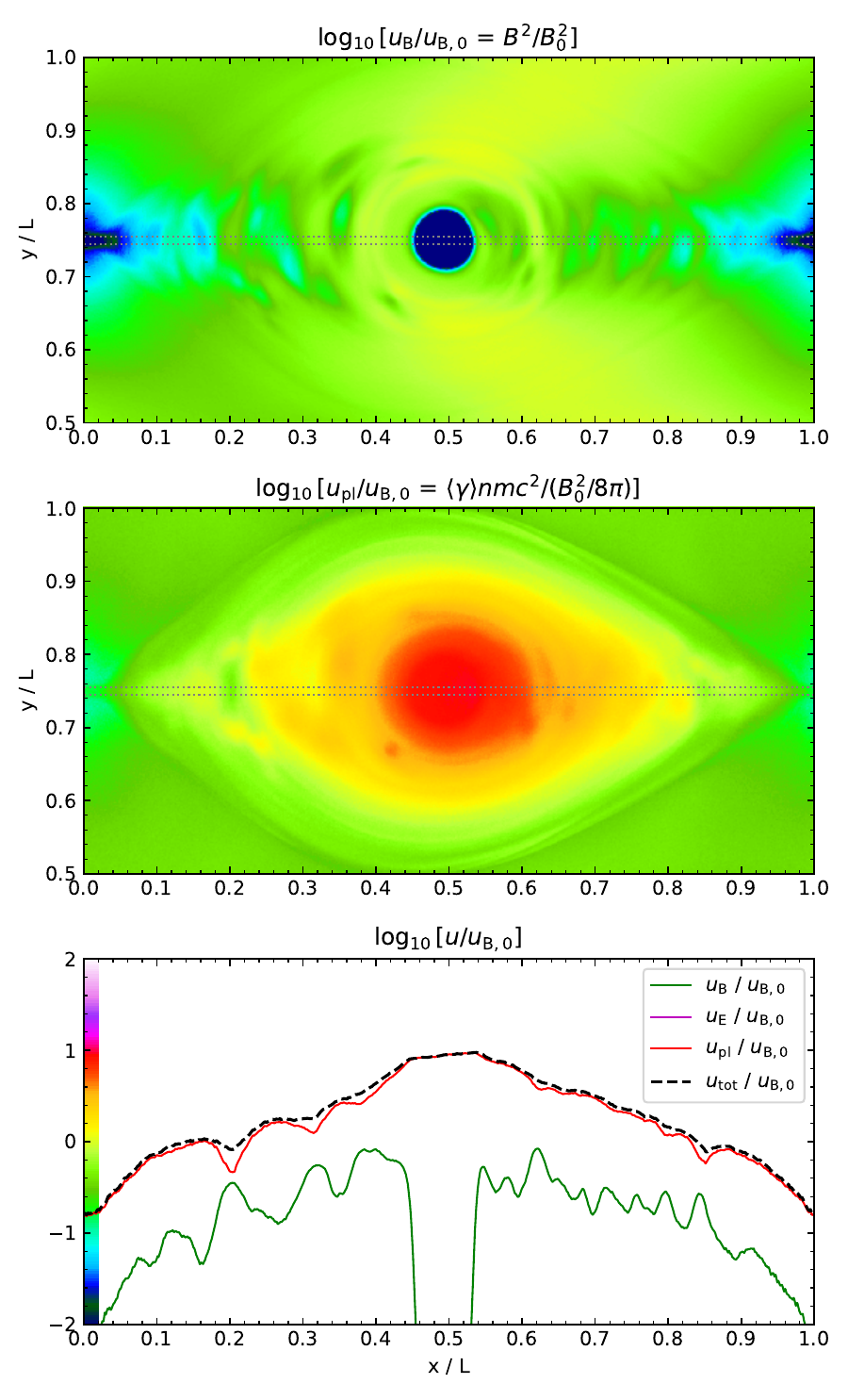}
\includegraphics[width=0.33\textwidth]{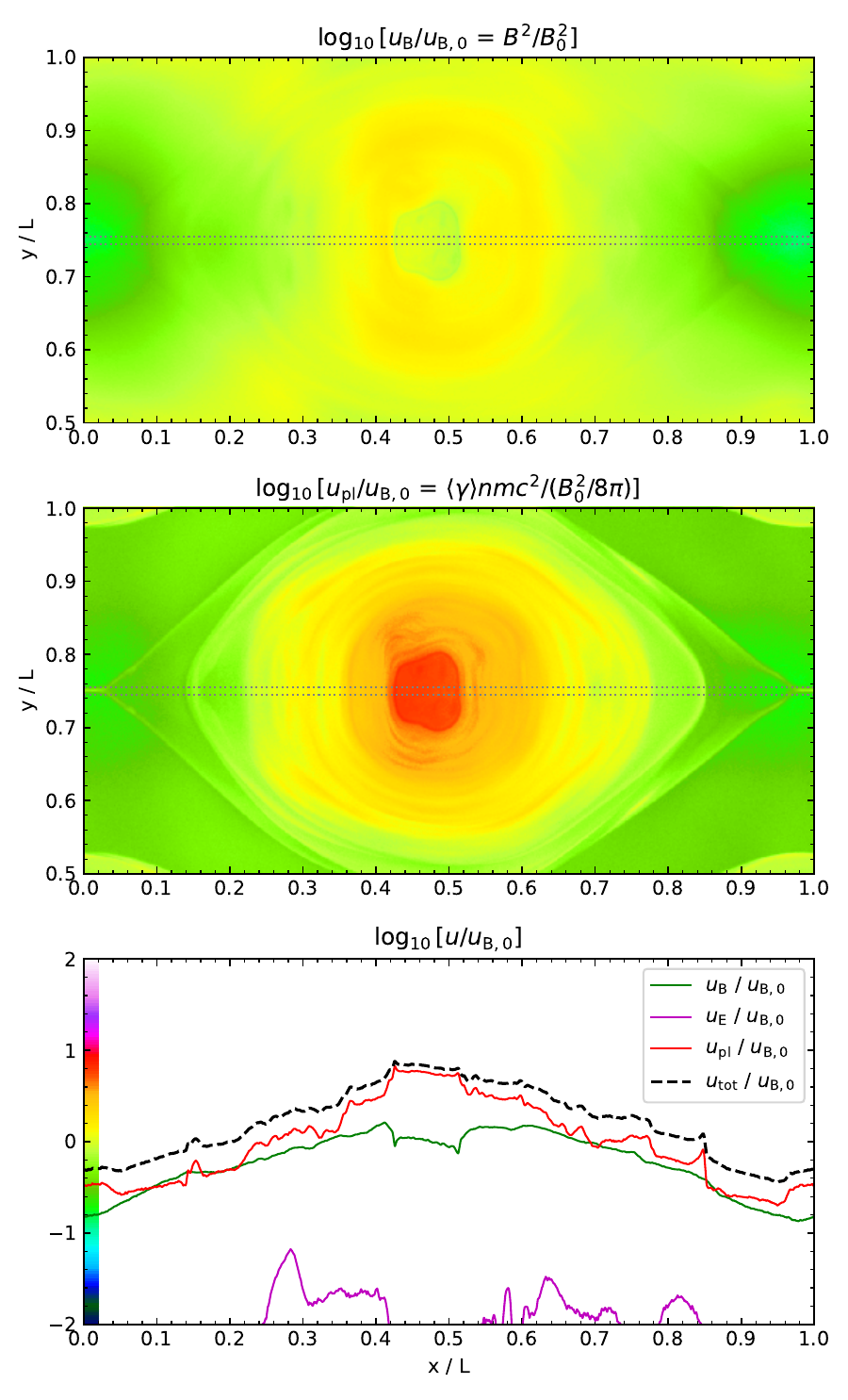}
\includegraphics[width=0.33\textwidth]{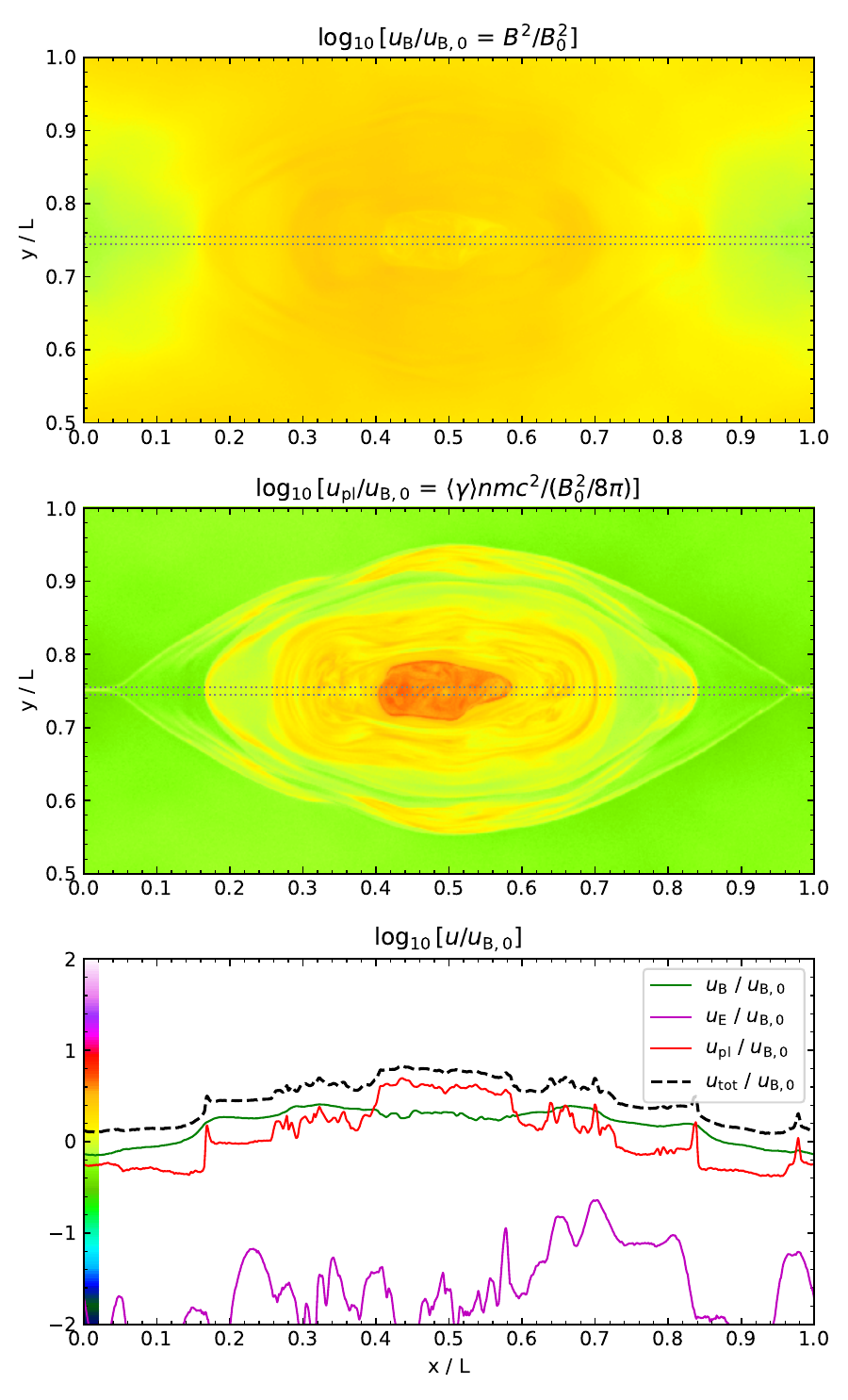}
\caption{
Maps $\log_{10}(u/u_{\rm B,0})(x,y)$ for magnetic energy density $u_{\rm B} = B^2/8\pi$ (upper panels) and plasma energy density $u_{\rm pl} = \left<\gamma\right> n m c^2$ (middle panels) for relaxed monster plasmoids at the end of each simulation.  In the lower panels, we compare 1D energy density profiles of $(u/u_{\rm B,0})(x,y_0)$ measured along the strip indicated in the above maps by the dashed gray lines.  A common color scale for all maps is referenced along the left axes in the lower panels.
Here we present the effect of guide field for $\sigma_0 = 10$ and $L/\rho_0 = 1800$. From the left, the columns show simulations: (1) L1800\_$\sigma$10, (2) L1800\_$\sigma$10\_Bg05, (3) L1800\_$\sigma$10\_Bg1.}
\label{fig_maps_relaxed_Bg}
\end{figure*}

\begin{figure*}
\includegraphics[width=0.245\textwidth]{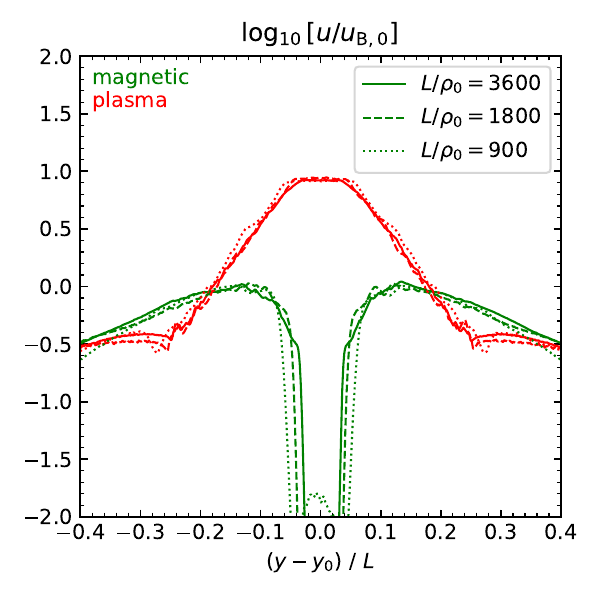}
\includegraphics[width=0.245\textwidth]{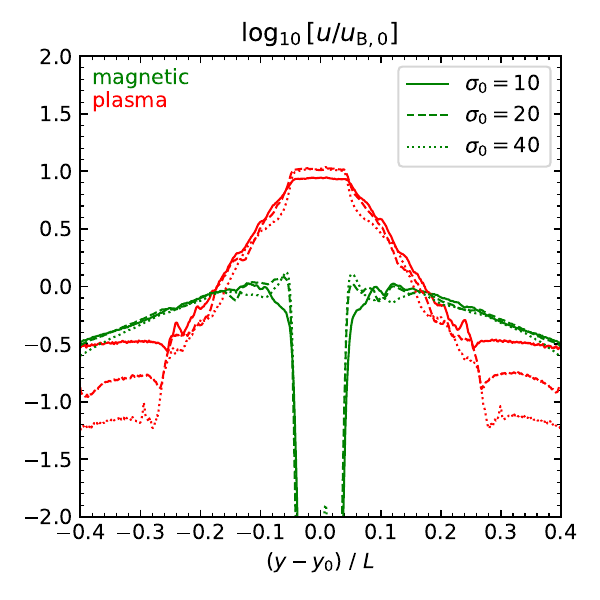}
\includegraphics[width=0.245\textwidth]{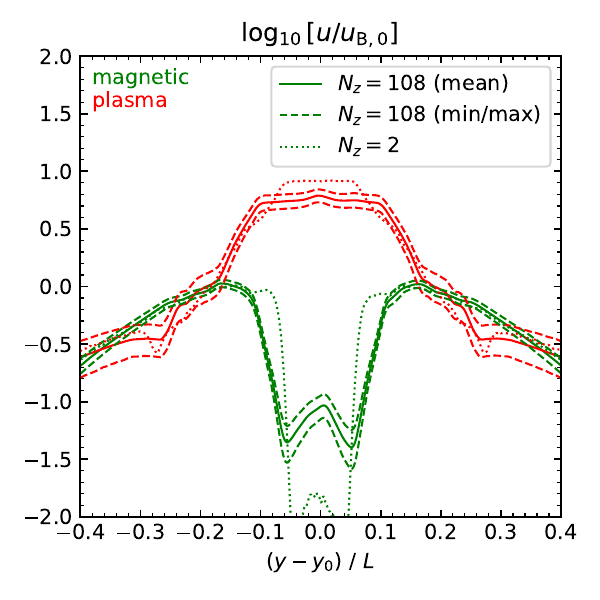}
\includegraphics[width=0.245\textwidth]{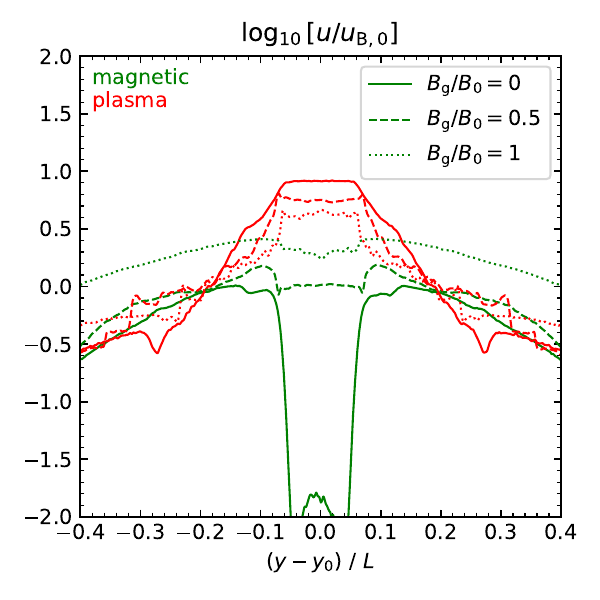}
\caption{Profiles of magnetic (green lines) and plasma (red lines) energy densities across (i.e., along the $y$ coordinate) relaxed monster plasmoids.
Panels from the left compare:
(1) different sizes $L/\rho_0$ of simulation domain (2D with $\sigma_0 = 10$, $B_z = 0$);
(2) different magnetisations $\sigma_0$ (2D with $L/\rho_0 = 1800$, $B_z = 0$);
(3) 3D and 2D domains (with $L/\rho_0 = 900$, $\sigma_0 = 10$, $B_z = 0$);
(4) different guide field strengths $B_{\rm g}/B_0$ (2D with $\sigma_0 = 10$, $L/\rho_0 = 1800$).
}
\label{fig_yprofs_relaxed}
\end{figure*}

\subsection{New simulations}
\label{sec_plasmoids_sims}

We used a modified version of the public PIC code {\tt Zeltron} \citep{Cer13} to perform effectively-2D and 3D simulations of relativistic magnetic reconnection with electron-positron pair plasma without radiation reaction in Cartesian coordinates $(x,y,z)$ with periodic boundaries.
The physical domain was $0 \le x \le L_x$, $0 \le y \le L_y$, $0 \le z \le L_z$ with $L_z < L_y = L_x$.
The numerical resolution was ${\rm d}x = L_x/N_x = L_y/N_y = L_z/N_z = \rho_0/2.56$ with the nominal gyroradius $\rho_0 = \Theta_{\rm e}m_{\rm e}c^2/eB_0$ with $m_{\rm e}$ the electron mass, $c$ the speed of light, $e$ the electric charge; $\Theta_{\rm e} = k_{\rm B}T_{\rm e}/m_{\rm e}c^2 = 1$ the relativistic temperature of initial Maxwell-J{\"u}ttner electron distribution with $k_{\rm B}$ the Boltzmann constant; with $B_0 = 1\;{\rm G}$ the nominal magnetic field strength.
We set two Harris layers: $B_x(y < y_{1/2}) = -B_0\tanh[(y-y_{1/4})/\delta]$ and $B_x(y \ge y_{1/2}) = B_0\tanh[(y-y_{3/4})/\delta]$ with $y_{1/2} = L_y/2$, $y_{1/4} = L_y/4$, $y_{3/4} = 3L_y/4$; the layer thickness $\delta = 2\rho_0/u_{\rm dr}$ is determined by the drift velocity $\beta_{\rm dr} = 0.3$ via $u_{\rm dr} \equiv \gamma_{\rm dr}\beta_{\rm dr} \equiv \beta_{\rm dr}/(1-\beta_{\rm dr}^2)^{1/2}$ \citep{2003ApJ...591..366K}.
The layers were supported by the pressure of drifting particle population of density $n_{\rm dr}(y < y_{1/2}) = n_{\rm dr,0} \cosh^{-2}[(y-y_{1/4})/\delta]$ (and likewise for $y > y_{1/2}$, centered at $y_{3/4}$) with $n_{\rm dr,0} = \gamma_{\rm dr} B_0^2/(8\pi\Theta_{\rm e}m_{\rm e}c^2)$.
A static background particle population of uniform density $n_{\rm bg}$ was added to achieve the desired background magnetisation $\sigma_0 = B_0^2/(4\pi n_{\rm bg}\Theta_{\rm e}m_{\rm e}c^2)$.
In some cases we added a guide magnetic field component of uniform strength $B_z$.
We did not use any initial perturbation.

\begin{table}
\caption{List of performed simulations with key parameters.}
\begin{tabular}{lrrrrr}
\hline\hline
label & $L_x/\rho_0$ & $N_x$ & $N_z$ & $\sigma_0$ & $B_{\rm g}/B_0$ \\
\hline
L3600\_$\sigma$10        & 3600 & 9216 &   2 & 10 & 0   \\ 
L1800\_$\sigma$10        & 1800 & 4608 &   2 & 10 & 0   \\ 
L1800\_$\sigma$20        & 1800 & 4608 &   2 & 20 & 0   \\ 
L1800\_$\sigma$40        & 1800 & 4608 &   2 & 40 & 0   \\ 
L1800\_$\sigma$10\_Bg0.5 & 1800 & 4608 &   2 & 10 & 0.5 \\ 
L1800\_$\sigma$10\_Bg1   & 1800 & 4608 &   2 & 10 & 1   \\ 
L1800\_$\sigma$10\_Lz14  & 1800 & 4608 &  36 & 10 & 0   \\ 
 L900\_$\sigma$10        &  900 & 2304 &   2 & 10 & 0   \\ 
 L900\_$\sigma$20        &  900 & 2304 &   2 & 20 & 0   \\ 
 L900\_$\sigma$40        &  900 & 2304 &   2 & 40 & 0   \\ 
 L900\_$\sigma$10\_Bg0.5 &  900 & 2304 &   2 & 10 & 0.5 \\ 
 L900\_$\sigma$10\_Bg1   &  900 & 2304 &   2 & 10 & 1   \\ 
 L900\_$\sigma$10\_Lz42  &  900 & 2304 & 108 & 10 & 0   \\ 
\hline\hline
\end{tabular}
\tablefoot{Simulations were performed in 3D domains with resolution of $N_x, N_x, N_z$. The domain length $L_x$ is reported in units of nominal gyroradius $\rho_0$. We report background magnetisation $\sigma_0$ and strength of guide magnetic field $B_{\rm g}$ in units of nominal field strength $B_0$.}
\label{tab_sims}
\end{table}

The list of performed simulations and their key parameters is presented in Table \ref{tab_sims}.
The cases for $N_z = 2$ are essentially 2D.
Simulations were performed for durations of at least $3 L_x/c$, over which they developed plasmoid chains that merged to one large relaxed plasmoid per layer.
The magnetic energy dissipation efficiency $\mathcal{E}_{\rm B}/\mathcal{E}_{\rm B,ini}$ was at least 50\% in all cases without guide field, almost 30\% by $5 L_x/c$ for $B_z/B_0 = 0.5$, and roughly 10\% by $4.5 L_x/c$ for $B_z/B_0 = 1$.

In simulations with guide-field, the total electric energy $\mathcal{E}_{\rm E}$ decays at a regular but slow rate of $\sim 20\%$ per $L/c$ (reflecting elastic oscillations of the plasmoid core), they were thus extended to durations of $t \sim 6 L/c$, by which point $\mathcal{E}_{\rm E} \sim 10^{-2}\mathcal{E}_{\rm B,0}$.
In contrast, simulations without guide field (for any considered $L/\rho_0$ or $\sigma_0$) achieved $\mathcal{E}_{\rm E} \sim 10^{-2.5}\mathcal{E}_{\rm B,0}$ by $t \lesssim 4 L/c$.

We calculated volume distributions $F_a(\mu) = {\rm d}F_a/{\rm d}\mu$ of parameter $\mu_a = \log_{10}(u_a/u_{\rm B,0})$ with $u_a$ the energy density of either magnetic field $u_{\rm B} = B^2/8\pi$ ($u_{\rm B,0} = B_0^2/8\pi$) or plasma $u_{\rm pl} = \left<\gamma\right>nm_{\rm e}c^2$ (including both electrons and positrons).
The distributions are normalised to unity, so that ${\rm d}\mu_a \sum_i F_a(\mu_{a,i}) = 1$.

\subsection{New results}
\label{sec_plasmoids_results}

Figure \ref{fig_voldist} compares the functions $f_a(\mu_a) = \log_{10}(F_a(\mu_a))$ averaged over entire durations of each simulation.
The peaks of the $F_{\rm B}$ distributions correspond to the initial background magnetic field strength $\mu_{\rm B,ini} = \log_{10}(1+B_z^2/B_0^2)$.
The peaks of the $F_{\rm pl}$ distributions correspond to the initial background plasma population ($\left<\gamma\right> \simeq 3.37$ for $\Theta_{\rm e} = 1$): $\mu_{\rm pl,bg} \simeq \log_{10}(6.7/\sigma_0)$, the initial drifting population contributes up to $\mu_{\rm pl,dr} \simeq \log_{10}(3.4\gamma_{\rm dr}^2) \simeq 0.57$.

High-energy-density tails of the distributions have complex structure that can be evaluated at different levels.
For the reference case of $\sigma_0 = 10$ without guide field, comparing the results for simulations {L1800\_$\sigma$10 and L3600\_$\sigma$10}, convergence for $F_{\rm B}$ was achieved to the level of $10^{-4}$, and for $F_{\rm pl}$ even down to $10^{-7}$.
The case of $\sigma_0 = 20$ is converged in $F_{\rm pl}$ down to $10^{-5}$.

\subsubsection{Relaxed monster plasmoids}

Monster plasmoids are very large ($\gtrsim 0.2 L$) and slow (${\rm v} \ll {\rm v}_{\rm A}$) plasmoids emerging from hierarchical mergers of plasmoids chains.
In simulations performed with periodic boundary conditions, they dominate the final states to which reconnecting plasma relaxes.

The parameter with the most qualitative impact on the structure of monster plasmoids is the strength of the guide field $B_{\rm g}$ relative to $B_0$.
Thus, Figure \ref{fig_maps_relaxed_Bg} presents energy density maps for relaxed monster plasmoids in the final states of 3 simulations for $B_{\rm g}/B_0 = 0, 0.5, 1$.

In the case of no guide field $B_{\rm g} = 0$ (L1800\_$\sigma$10), the plasma energy density has a smooth structure, while magnetic energy density is rippled, especially along the $x$ axis.
The plasmoid is dominated by plasma energy density with $u_{\rm pl}/u_{\rm B} \sim 10$, which peaks at $\mu_{\rm pl} \simeq 0.9$ across the {central} unmagnetised core {for $0.45 < x/L < 0.54$}; in the $f_{\rm pl}$ distribution {(Figure \ref{fig_voldist}, right panel, red solid line)} this corresponds to a break at the $-1$ level.
The magnetic energy density peaks at $\mu_{\rm B} \simeq -0.05$ (just below the initial peak of $f_{\rm B}$; {Figure \ref{fig_voldist}, left panel, red solid line}) along a ring structure crossing the $y/L = 0.75$ plane {at $x/L \simeq 0.39$ and $0.62$}.
Hence, a relaxed monster plasmoid achieves only a minor compression of $u_{\rm pl}$, and no compression of $u_{\rm B}$.

In the case of strong guide field $B_{\rm g} = B_0$ (L1800\_$\sigma$10\_Bg1),
the plasmoid is characterised by roughly uniform magnetic energy density with $\mu_{\rm B} \sim 0.3$ (corresponding to the peak of $f_{\rm B}$; {Figure \ref{fig_voldist}, left panel, black solid line}).
The plasma energy density shows complex substructures with high-density arcs tracing the magnetic field lines.
Only the plasmoid core ($0.4 < x/L < 0.59$) is dominated by the plasma with $\mu_{\rm pl} \simeq 0.6$ (corresponding to a bump in $f_{\rm pl}$ at the $-1.1$ level; {Figure \ref{fig_voldist}, right panel, black solid line}) and hence $u_{\rm pl}/u_{\rm B} \sim 2$, the plasmoid layer is magnetically dominated with $u_{\rm B}/u_{\rm pl} \sim 1.8$.

The case of moderate guide field $B_{\rm g} = B_0/2$ (L1800\_$\sigma$10\_Bg05) shows intermediate structure, which is qualitatively closer to the case of strong guide field.

\begin{figure*}
\includegraphics[width=0.33\textwidth]{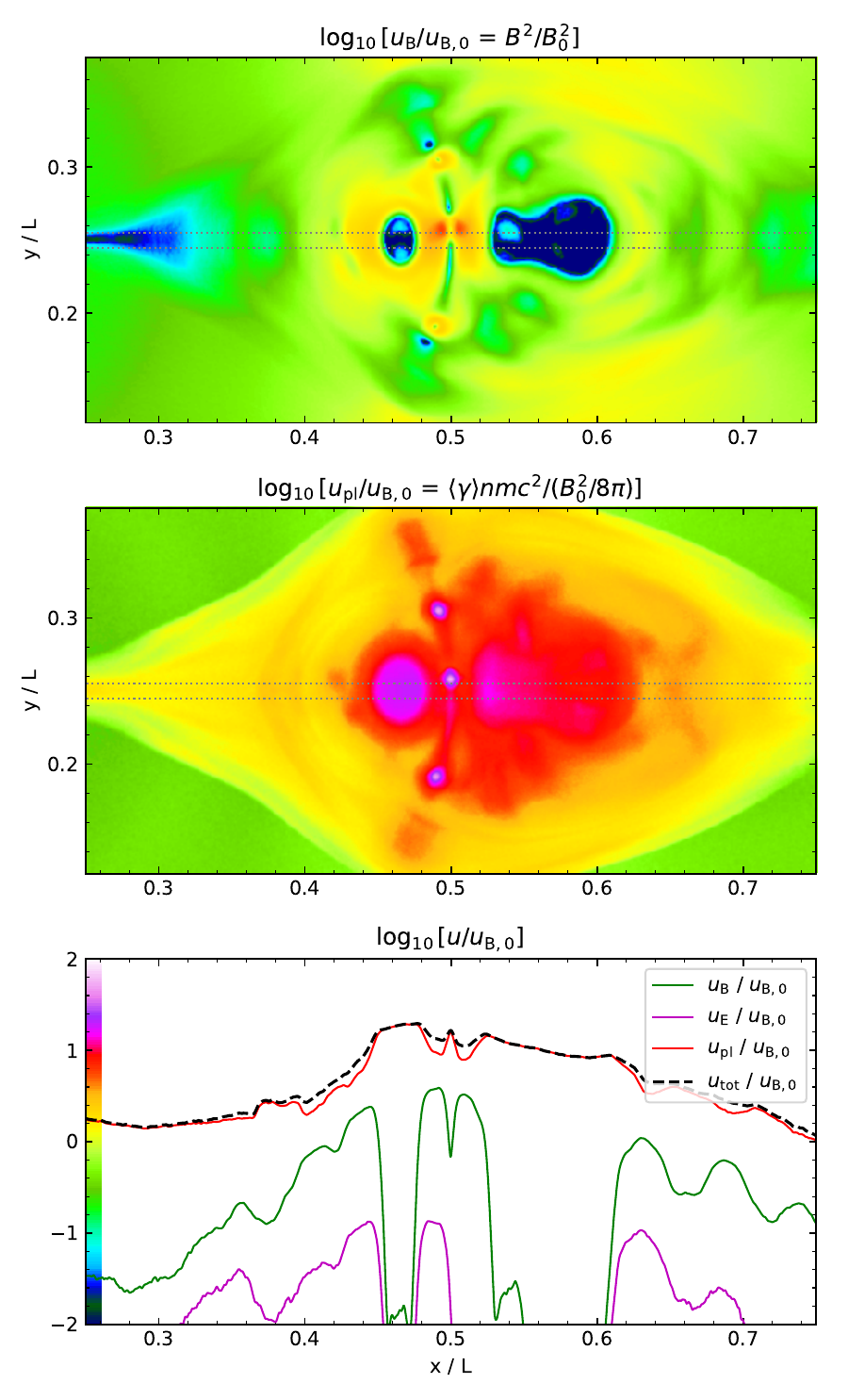}
\includegraphics[width=0.33\textwidth]{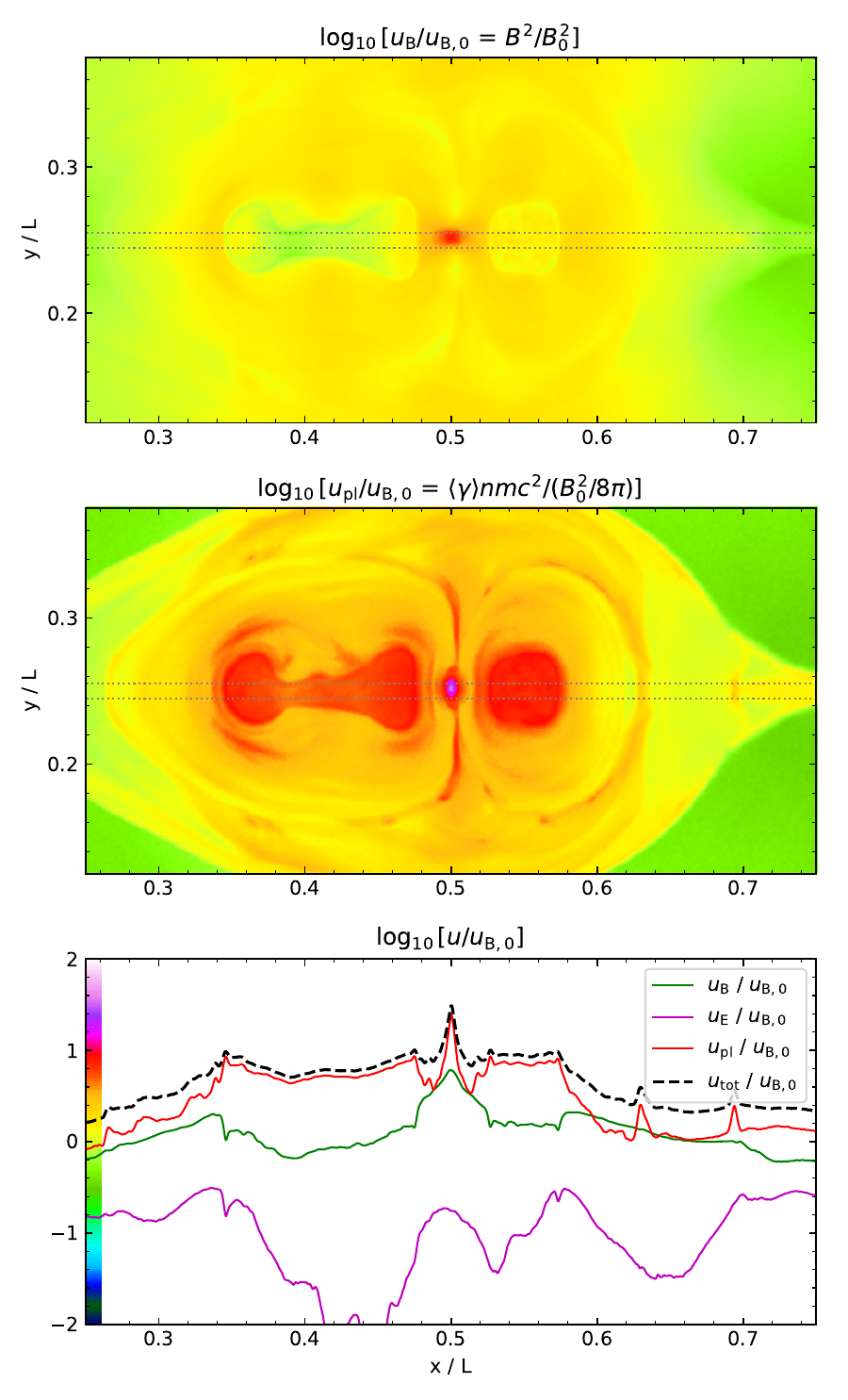}
\includegraphics[width=0.33\textwidth]{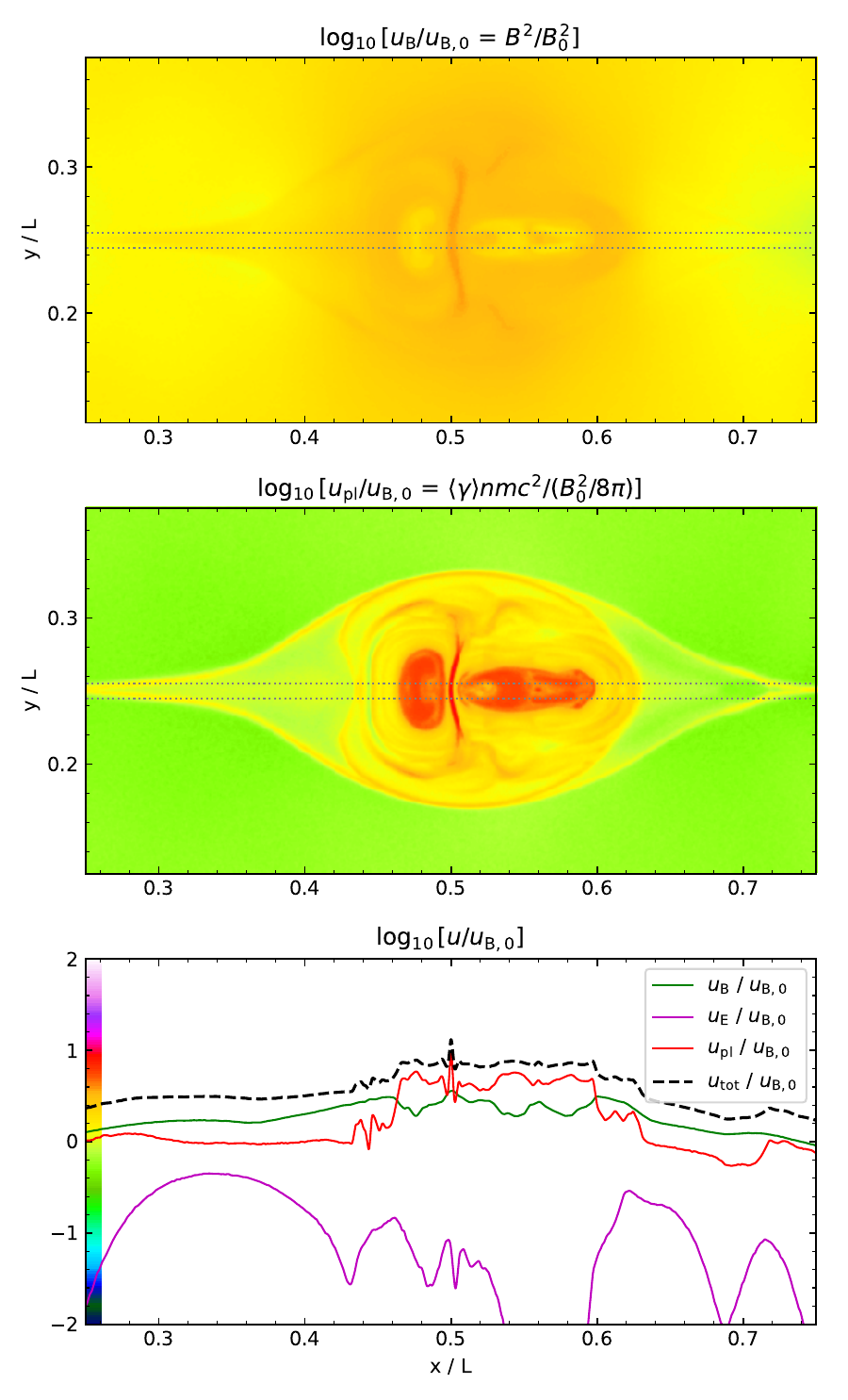}
\caption{Same as Figure \ref{fig_maps_relaxed_Bg}, but for plasmoid mergers that maximize the plasma energy density $u_{\rm pl}$ for the same set of simulations: (1) L1800\_$\sigma$10, (2) L1800\_$\sigma$10\_Bg05, (3) L1800\_$\sigma$10\_Bg1.}
\label{fig_maps_mergers}
\end{figure*}

Figure \ref{fig_yprofs_relaxed} compares the $y$-profiles of magnetic and plasma energy densities measured along the $y$ axis of relaxed monster plasmoids.
The first (from the left) panel compares such profiles for 3 sizes of simulation domain $L/\rho_0 = 900, 1800, 3600$ (with no guide field, $\sigma_0 = 10$).
The general characteristic of these profiles (specific to the cases without guide field) is a very sharp transition between the plasmoid core ($0.455 < y/L < 0.54$, $u_{\rm B} < u_{\rm B,0}/100$, uniform $u_{\rm pl}$) and the plasmoid layer, supported by a sharp ring structure of electric current density $j_z$.
Across the plasmoid core, the plasma energy density is roughly uniform at the level of $u_{\rm pl} \simeq 8.5 u_{\rm B,0}$.
It clearly dominates the magnetic energy density in the plasmoid layer, peaking at the level of $u_{\rm B,peak} \simeq u_{\rm B,0}$, while remaining in pressure balance -- a key factor is that our choice of relativistic plasma temperature $\Theta_{\rm e} = 1$ implies a relation\footnote{For isotropic Maxwell-J{\"u}ttner distribution, $u_{\rm pl}/P_{\rm pl}$ equals the mean particle Lorentz factor $\left<\gamma_{\rm e}\right> \simeq 3\Theta_{\rm e} + K_1(1/\Theta_{\rm e})/K_2(1/\Theta_{\rm e})$, with $K_n$ the modified Bessel function of the second kind.} between plasma energy density and pressure $u_{\rm pl}/P_{\rm pl} \simeq 3.4$.
Across the plasmoid layer, the plasma energy density decreases exponentially like $u_{\rm pl} \propto 10^{-|y-y_0|/(L/7)}$, bringing the magnetic field towards the force-free equilibrium $u_{\rm B} \propto 1/|y-y_0|^2$ for $|y-y_0|/L > 0.28$.
With increasing $L/\rho_0$, the plasmoid core becomes smaller in units of $L$, and also larger in units of $\rho_0$, it scales approximately as $\propto (L/\rho_0)^{1/2}$.
However, this has little effect on the profiles of $u_{\rm pl}$, both in the core and across the layer.

The second panel of Figure \ref{fig_yprofs_relaxed} compares the profiles of $u_{\rm B}$ and $u_{\rm pl}$ for 3 values of initial background magnetisation $\sigma_0 = 10, 20, 40$ (with no guide field, $L/\rho_0 = 1800$).
The effect of $\sigma_0$ on the core and the layer of the plasmoid is rather minor (a clear effect of $\sigma_0$ can be seen for $y/L < 0.25$ and $y/L > 0.75$, where background plasma still dominates).
For $\sigma_0 = 10$, the magnetic boundary between the core and the layer is less sharp, which is balanced by a slightly lower $u_{\rm pl}$ across the core.

The third panel of Figure \ref{fig_yprofs_relaxed} compares the profiles of $u_{\rm B}$ and $u_{\rm pl}$ between an essentially 2D simulation (with $N_z = 2$) and a 3D simulation with $L_z/\rho_0 \simeq 42$ (with $N_z = 108$), other key parameters being exactly the same ($L_x = L_y = 900\rho_0$, $\sigma_0 = 10$, $B_{\rm g} = 0$).
Profiles in the 3D case were calculated using different statistics along the $z$ coordinate: minimum, mean, maximum.
The resulting monster plasmoid showed a broader core with gradual boundaries -- a plateau of $u_{\rm pl} \sim 6 u_{\rm B,0}$ with a total width of $\Delta y \sim 0.2L$, and a base of $u_{\rm B} \sim 0.07 u_{\rm B,0}$ with a total width of $\Delta y \sim 0.1L$.
The plasmoid layer is consistent between 3D and 2D cases.

The last panel of Figure \ref{fig_yprofs_relaxed} compares the profiles of $u_{\rm B}$ and $u_{\rm pl}$ for 3 values of guide field strength $B_{\rm g}/B_0 = 0, 0.5, 1$ (with $L/\rho_0 = 1800$, $\sigma_0 = 10$).
In the presence of guide field, both profiles appear less regular, which reflects a much slower relaxation of these plasma configurations.
Nevertheless, one can notice that the profiles of $u_{\rm B}$ are flatter, with plasmoid cores filled with uniform $u_{\rm B}$ at the levels of $\sim u_{\rm B,0}$ for $B_{\rm g} = B_0/2$, and $\sim 2u_{\rm B,0}$ for $B_{\rm g} = B_0$.
Still, the plasmoid cores are dominated by $u_{\rm pl}$, in the $B_{\rm g} = B_0$ case by factor $\sim 2$.
The plasmoid layer is roughly in equipartition for $B_{\rm g} = B_0/2$ and magnetically dominated for $B_{\rm g} = B_0$.

\begin{figure*}
\includegraphics[width=\textwidth]{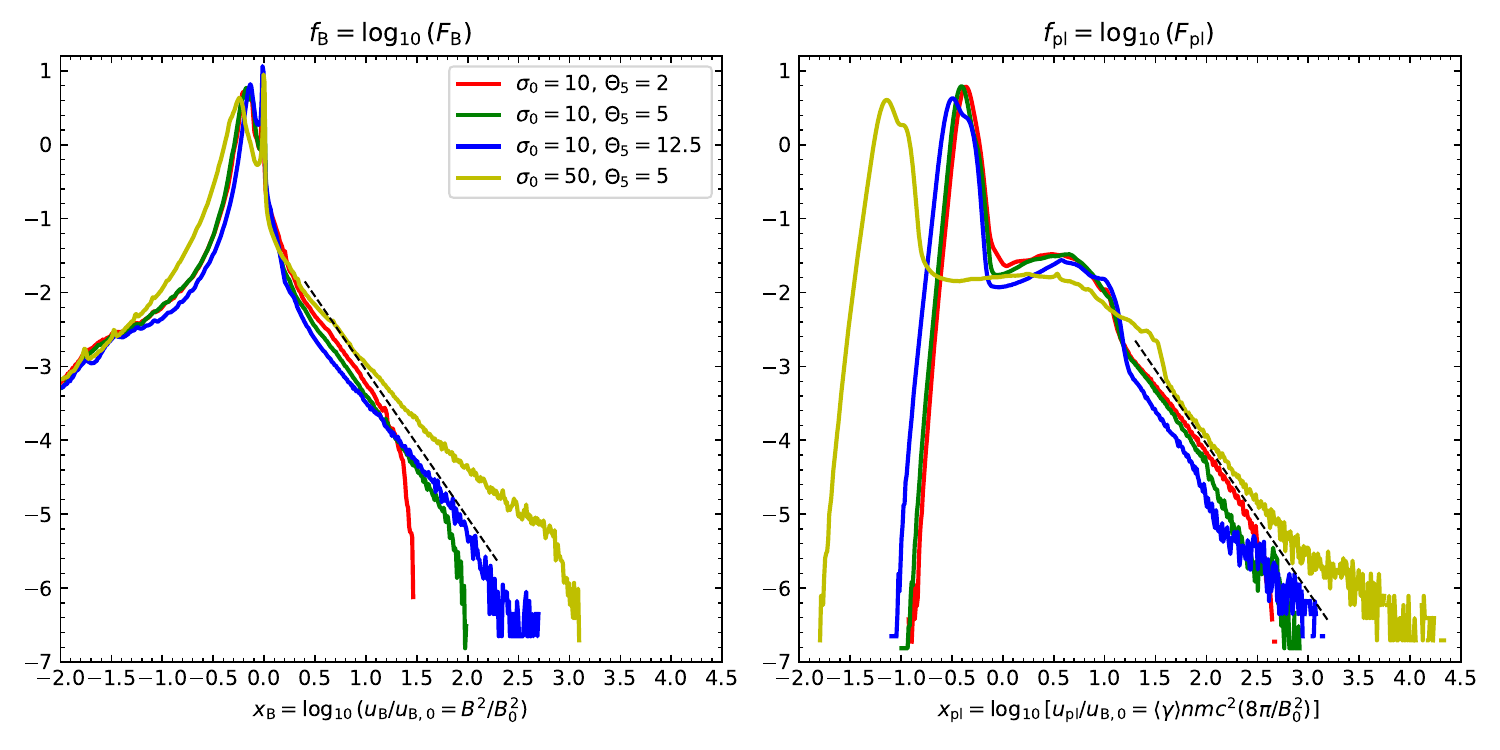}
\caption{Same as Figure \ref{fig_voldist}, but for the open-boundary simulations with synchrotron cooling first presented in \cite{2020MNRAS.497.1365O}.
The black dashed lines indicate power laws of index $-5/3$.}
\label{fig_voldist_openb}
\end{figure*}

\subsubsection{Plasmoid mergers}

Figure \ref{fig_maps_mergers} shows energy density maps for merging plasmoids in the same three simulations {as in Figure \ref{fig_maps_relaxed_Bg}}.

In the reference case {L1800\_$\sigma$10}, the entire system of merging plasmoids including the secondary current layer is dominated by plasma energy density {reaching $\mu_{\rm pl} \sim 1.3$} (corresponding {to $f_{\rm pl} \simeq -3$} in the high-density tail; {Figure \ref{fig_voldist}, right panel, red solid line}).
The magnetic energy density {reaches $\mu_{\rm B} \sim 0.6$} (corresponding {to $f_{\rm B} \simeq -3.7$}; {Figure \ref{fig_voldist}, left panel, red solid line}) in the vicinity of the current layer (possibly related to the formation of a minor secondary plasmoid), the electric energy density {reaches $\mu_{\rm E} \sim -0.9$} in the same region.
Compared to the relaxed monster plasmoid, compression of $u_{\rm pl}$ is stronger by {factor $\sim 2.5$}, but compression of $u_{\rm B}$ is stronger by {factor $\sim 4.5$}.

In the case of strong guide field $B_{\rm g} = B_0$ (L1800\_$\sigma$10\_Bg1),
we show a merger that forms a secondary current layer where plasma energy density reaches $\mu_{\rm pl} \simeq 0.95$ (corresponding to $f_{\rm pl} \simeq -4$; {Figure \ref{fig_voldist}, right panel, black solid line}; higher by factor $\simeq 2.2$ than in the monster plasmoid)
and magnetic energy density reaches $\mu_{\rm B} \simeq 0.55$ (corresponding to $f_{\rm B} \simeq -2$; {Figure \ref{fig_voldist}, left panel, black solid line}; higher by {factor $\simeq 1.8$} than in the monster plasmoid).

The case of moderate guide field $B_{\rm g} = B_0/2$ (L1800\_$\sigma$10\_Bg05) shows {plasma} energy densities {comparable} to the case of no guide field.

\subsubsection{Structures in the volume distributions}

With this analysis, we can identify the main structures in the volume distributions $f_{\rm pl}(\mu_{\rm pl})$ and $f_{\rm B}(\mu_{\rm B})$ presented in Figure \ref{fig_voldist}.

The plateau in the $f_{\rm pl}$ distribution corresponds to relaxed plasmoids.
In the absence of guide field, it extends to $\mu_{\rm pl} \simeq 1$ or $u_{\rm pl}/u_{\rm B,0} \sim 10$, only weakly depending on background magnetisation $\sigma_0$.
Compared with the initial drifting particle population, this means a modest compression by factor $\sim 2.7$.
A guide field of $B_z = B_0$ reduces this compression by factor $\sim 1.8$.
Such plateau is absent in the $f_{\rm B}$ distribution --- relaxed plasmoids do not amplify $u_{\rm B}$.

The steep high-density tail of the $f_{\rm pl}$ and $f_{\rm B}$ distributions, extending to the level of $-4$, corresponds to plasmoid mergers.
Considering the tail of the $f_{\rm pl}$ distribution at the level of $-4$,
in our reference case ($\sigma_0 = 10$ in the absence of guide field) it reaches $\mu_{\rm pl} \simeq 1.5$.
Compared with the initial drifting particle population, this means compression by factor $\sim 8.5$.
A higher magnetisation of $\sigma_0 = 40$ enhances this compression by factor $\sim 1.6$, while a guide field of $B_z = B_0$ reduces it by factor $\sim 3.5$.
Considering the tail of $f_{\rm B}$ at the same level of $-4$, for $\sigma_0 = 10$ in the absence of guide field it reaches $\mu_{\rm B} \simeq 0.75$ (compression of $u_{\rm B,0}$ by factor $\sim 5.5$), for $\sigma_0 = 40$ compression increases by factor $\sim 1.4$, while for $B_z = B_0$ it decreases by factor $\sim 1.25$.

The second high-density tail of the $f_{\rm pl}$ distribution, extending below the $-4$ level, can be identified as secondary plasmoids.
Considering the level of $-6$, in the reference case it reaches at least $\mu_{\rm pl} \simeq 2.1$ (factor $\sim 4$ higher than for primary merging plasmoids).
In the cases of $\sigma_0 = 20, 40$, it reaches $\mu_{\rm pl} \simeq 2.4$ (factor $\sim 2$ higher), and in the guide field case $B_z = B_0$ it reaches $\mu_{\rm pl} \simeq 1.35$ (factor $\sim 5.5$ lower).

\subsection{Comparison with open-boundary radiative simulations}
\label{sec_plasmoids_compare}

Figure \ref{fig_voldist_openb} presents volume distributions of magnetic and plasma energy densities for simulations of relativistic reconnection initiated from a single Harris current layer using open boundary conditions and including synchrotron cooling, first presented in \cite{2020MNRAS.497.1365O}.
Open boundary conditions were applied across both ends of the current layer ($x = 0$ and $x = L_x$), which allowed for free escape of plasmoids and prevented formation of a single slow (due to cancellation of $x$-momentum) monster plasmoid.
The background plasma was extended to a larger volume ($L_z = 4 L_x$), largely preserving it in pristine condition throughout the simulation time, we thus find stronger dominance of background plasma at $u_{\rm B},u_{\rm pl} < u_{\rm B,0}$ ($x_{\rm B},x_{\rm pl} < 0$).
In the $F_{\rm pl}$ distribution, a bump for $0 < x_{\rm pl} < 1.1$ corresponding to large plasmoids is still prominent.
A major difference from periodic boundary simulations is the presence of extended power-law tails in both $F_{\rm B}$ and $F_{\rm pl}$ distributions.
In the high-magnetisation case ($\sigma_0 = 50$), the tail of $F_{\rm B}$ extends over $0.3 < x_{\rm B} < 2.8$, and the tail of $F_{\rm pl}$ extends over $1.6 < x_{\rm pl} < 3.7$.
The tails are less extended in the low-magnetisation cases ($\sigma_0 = 10$), and especially the cut-off of $F_{\rm B}$ is sensitive to radiative cooling efficiency, shifting to $x_{\rm B} \simeq 1.2$ for $\Theta = 2\times 10^5$ (slow cooling).
The power-law index is approximately $p \simeq -5/3$ for both $F_{\rm B}$ and $F_{\rm pl}$, although for $\sigma_0 = 50$ it hardens towards the highest $x_{\rm B},x_{\rm pl}$ to $p \simeq -1$, and for $\sigma_0 = 10$ it softens to $p \simeq -2$.

\section{Relativistic jets}
\label{sec_jets}

Relativistic jets are launched from rotating magnetospheres with large fluxes of poloidal magnetic field \citep{1976MNRAS.176..465B}.
The twist of poloidal magnetic field generates toroidal magnetic component and perpendicular electric field, combining into outflowing electromagnetic momentum -- the Poynting flux.
In the process of \cite{1977MNRAS.179..433B}, an electromagnetic version of the Penrose process \citep{2014PhRvD..89b4041L}, torque is exerted on the black hole (BH), converting its rotational energy to the energy of toroidal magnetic field and associated electric field.

A magnetosphere connected to the BH horizon achieves highly relativistic magnetisation ($\sigma = b^2/w \gg 1$) and force-free condition (negligible inertia $\rho c^2$, pressure $P$, enthalpy $w$; sufficient charge density $\rho_{\rm e}$, current density $\vec{j}$) by unloading its plasma onto the BH.
This results in a sharply edged cavity of low plasma density, which (in combination with relatively uniform magnetic enthalpy density $b^2$) corresponds with the $\sigma > 1$ magnetosphere \citep[e.g.,][]{2024A&A...692A..37N}.

An outflowing BH magnetosphere is gradually collimated by external plasma pressure and accelerated by pressure of the toroidal magnetic field \citep{1989ASSL..156..129C}, converting its electromagnetic energy into kinetic energy of the plasma, according to the conservation of Michel parameter (specific energy) $\mu = \Gamma(1+\sigma_\phi')$ \citep{1969ApJ...158..727M} -- magnetisation $\sigma_\phi' = B_\phi'^2/(4\pi w')$, based on the co-moving toroidal magnetic field $B_\phi'$ and relativistic enthalpy density $w'$, is converted to the bulk Lorentz factor $\Gamma$.
The efficiency of such acceleration depends critically on the geometry of poloidal magnetic field lines -- it is particularly low in the spherical case \citep{1994ApJ...426..269B}, but can be very high in the parabolic case \citep{2006MNRAS.367..375B}.
In this acceleration-collimation zone (ACZ), extending to distances $r ~ (10^3 - 10^5) M_{\rm BH}$ \citep{2007MNRAS.380...51K,2009MNRAS.394.1182K}, the jet is characterised by high magnetisation and largely ordered magnetic fields.

The boundary of a relativistic jet can be defined by trans-relativistic 4-velocity $u = \Gamma v/c = 1$, which corresponds to the Lorentz factor $\Gamma = (u^2+1)^{1/2} = \sqrt{2}$.
Within the jet is the density cavity, which in the ACZ can be identified by $\sigma = 1$ (beyond the ACZ, $\sigma < 1$ across the entire jet; whether the cavity remains well defined depends on the efficiency of plasma mixing processes).
Recent numerical works \citep[e.g.,][]{2018A&A...612A..34D,2024MNRAS.533..254S} describe the inner $\sigma > 1$ region as the jet spine, and the outer $\sigma < 1$ region as the jet sheath; the original spine-sheath structure has been proposed by \cite{2005A&A...432..401G}.

To describe the jet structure, we will use spherical coordinates $(r,\theta,\phi)$ and cylindrical radius $R = r\sin\theta$.
Gradual collimation in the ACZ means that the shape of the jet spine $\theta_{\rm sp}(r)$ can be described as paraboloidal; beyond the ACZ it transitions to conical -- this has been confirmed by direct measurements of the innermost jets in nearby AGNs with the very-long-baseline interferometry (VLBI) technique \citep{2012ApJ...745L..28A,2020MNRAS.495.3576K}.
A flat or slowly declining profile of external/sheath pressure may even lead to recollimation of the jet spine \citep{2007ApJ...671..678B,2015MNRAS.452.1089P}.


\subsection{Lateral structure of jet magnetic fields}

Relativistic jets can be robustly described by stationary and axisymmetric models.
In such models, one considers two main components of the magnetic and velocity fields that are closely interrelated.
The poloidal plasma velocity is parallel to the poloidal magnetic field, hence poloidal plasma streamlines $\theta_{\rm s}(r)$ coincide with the poloidal field lines.
The relation between poloidal and toroidal magnetic fields is governed by the trans-field momentum (Grad-Shafranov) equation \citep{2010mfca.book.....B}.
Detailed solutions presented in the literature \citep[e.g.,][]{1993A&A...274..699A,2009ApJ...698.1570L,2010MNRAS.402..353L,2023MNRAS.524.4012B} show several characteristic features.
One can start by describing the poloidal component, persistently anchored to the central engine; then discuss the toroidal component as a train of giant electromagnetic waves propagating along the poloidal streamlines.

Close to the black hole, the magnetosphere is uncollimated, hence the streamlines are roughly spherical, $\theta_{\rm s}(r) \simeq \theta_{\rm s}(r_{\rm H})$.
In the initial stage of uniform (equilibrium) collimation \citep{2010MNRAS.402..353L}, the entire jet remains in causal contact, and the streamlines are able to adapt to the jet spine boundary $\theta_{\rm sp}(r)$, to the tension of toroidal magnetic field, and to the centrifugal forces due to both azimuthal rotation and poloidal curvature \citep{2009ApJ...697.1681N}.
In the subsequent stage of differential (non-equilibrium) collimation, beyond the fast magnetosonic surface, the inner layer of poloidal field lines bunches towards the jet symmetry axis \citep{2009ApJ...699.1789T,2019MNRAS.490.2200C}.
An important consequence is that the strength of poloidal magnetic field $B_{\rm p}$ evolves very differently along individual field lines.

Close to the jet symmetry axis, magneto-hydro-dynamic (MHD) models predict a narrow cylindrical structure called the {jet core} \citep[e.g.,][]{1993A&A...274..699A,2009MNRAS.397.1486B,2009ApJ...698.1570L,2023MNRAS.524.4012B}.
Here, the poloidal magnetic field lines are able, more or less, to co-rotate at the rate imposed by the central engine (this corresponds roughly to the light cylinder of pulsars).
The characteristic radius of the jet core is $R_{\rm c} \sim u_{\rm c}/\Omega_{\rm B}$, where $u_{\rm c} = \Gamma_{\rm c}\beta_{\rm c} \sim 1$ is the minimal 4-velocity of bulk motion along the core\footnote{For a Kerr BH of mass $M_{\rm BH}$ and spin $a$ in natural units $G = c = 1$, the outer horizon radius is $r_{\rm H}/M_{\rm BH} = 1 + (1-a^2)^{1/2}$, and the angular velocity is $\Omega_{\rm H} = a/(2r_{\rm H})$.  The magnetosphere rotates with an angular velocity of $\Omega_{\rm B} \simeq \Omega_{\rm H}/2$ \citep{2001MNRAS.326L..41K}.}.
For $a \simeq 0.9$, this means $R_{\rm c} \simeq 6M_{\rm BH}$, effectively imposing the BH length scale over great distances.
A strictly cylindrical jet core would maintain constant $B_{\rm p}$, support only a weak toroidal field (hence little Poynting flux) and inefficient bulk acceleration with $\Gamma_{\rm j} \simeq B_\phi/B_{\rm p} \simeq \Omega_{\rm B}R$ \citep{2009ApJ...698.1570L}.

Outside the jet core ($R > R_{\rm c}$), differential collimation results in poloidal field lines diverging faster than in the spherical case, resulting in $B_{\rm p}$ declining steeper than $1/r^2$, enhancing the acceleration efficiency due to the magnetic nozzle effect \citep{1994ApJ...426..269B}.
Efficient conversion of magnetic to kinetic energy still requires maintaining equilibrium with the external pressure \citep{2010MNRAS.402..353L}.
On the other hand, a steeply decreasing external pressure (e.g., when a gamma-ray burst jet breaks out of a collapsar) enables additional acceleration of the outermost jet layers via rarefaction mechanism \citep{2010NewA...15..749T,2010MNRAS.407...17K}.

Toroidal magnetic field is induced by rotation of poloidal field lines at the jet base, where centrifugal force wins over BH gravity and dominates the magnetic tension.
BH magnetospheres do not rotate rigidly, and they inject toroidal field over a broad range of streamlines, well away from the jet core.
The relative strengths of toroidal and poloidal components define the {magnetic pitch} $\mathcal{P} = RB^r/B^\phi$ \citep{2000A&A...355..818A,2019ApJ...884...39B}.
The initial stage of jet collimation is characterised by roughly uniform $\mathcal{P}$.
At the jet core radius $R_{\rm c}$ one has $B^\phi \simeq B^r$, hence $\mathcal{P} \simeq R_{\rm c}$ across the jet.
This implies that radial profiles of toroidal field strength $B^\phi(R)$ peak well outside the jet core, and hence $R_{\rm c}$ is not their characteristic lateral scale.

\citep{2019MNRAS.490.2200C} performed axisymmetric general-relativistic MHD (GRMHD) simulations of strongly collimated relativistic jets extending to distances beyond $10^5 r_{\rm g}$.
They demonstrated strong bunching of poloidal field lines \citep{2009ApJ...699.1789T} leading to formation of jet core of radius $R_{\rm c} \simeq 4 R_{\rm L}$ \citep{2023MNRAS.524.4012B} (their $a = 15/16$ implies $\Omega_{\rm B} \simeq \Omega_{\rm H}/2 \simeq 0.17$, hence $R_{\rm L} \simeq 5.7 R_{\rm g}$).
On one hand, bunching of poloidal field lines is a consequence of magnetic tension of the toroidal field;
on the other hand, bunched poloidal field lines transmit the toroidal field parallel to the jet core.
The toroidal magnetic energy gradually converts to kinetic energy of the plasma, resulting in high bulk Lorentz factors in the vicinity of the jet core, which was also reported by \cite{2019MNRAS.490.2200C}.
{At the same time,} the outer jet regions are loaded with denser proton plasma from the sheath region, apparently due to the interchange (magnetic Rayleigh-Taylor) instability\footnote{\cite{2019MNRAS.490.2200C} call this a pinch instability, with `pinch' referring to longitudinal $\propto\exp(ikr)$ perturbations of the spine-sheath boundary; `pinch' usually refers to the $m=0$ mode of current-driven perturbations $\propto \exp(im\phi)$.}.
Mass loading via instabilities is expected to develop a highly inhomogeneous filamentary structure; whether such filaments can reach the jet core is unknown.
It has been suggested that filamentary mass loading would result in highly inhomogeneous jet magnetisation with $\sigma_{\rm max} \gtrsim 10^3$ matching the maximum particle energies inferred from non-thermal emission of blazars \citep{2016Galax...4...28N}.
On average, mass loading should reduce not only the bulk Lorentz factor $\Gamma$, but also the Michel parameter $\mu = \Gamma(1+\sigma)$ (maximum achievable $\Gamma$, hence the potential for further acceleration).
The overall effect of poloidal field bunching is thus concentration of the jet energy flux along the jet core.

The poloidal pressure of the toroidal magnetic field is the main force accelerating the jet to relativistic speeds.
Non-uniform structure of $B^\phi(\theta)$ results in non-uniform structure of $\Gamma(\theta)$.
The acceleration efficiency depends also on the geometry of jet streamlines (rate of velocity divergence).
Thus, acceleration should be less efficient in the jet core, and particularly along the jet axis.
The toroidal magnetic field is the main carrier of the poloidal Poynting flux, which converts to plasma kinetic energy density along each streamline.


\begin{figure*}
\centering
\includegraphics[width=0.7\textwidth]{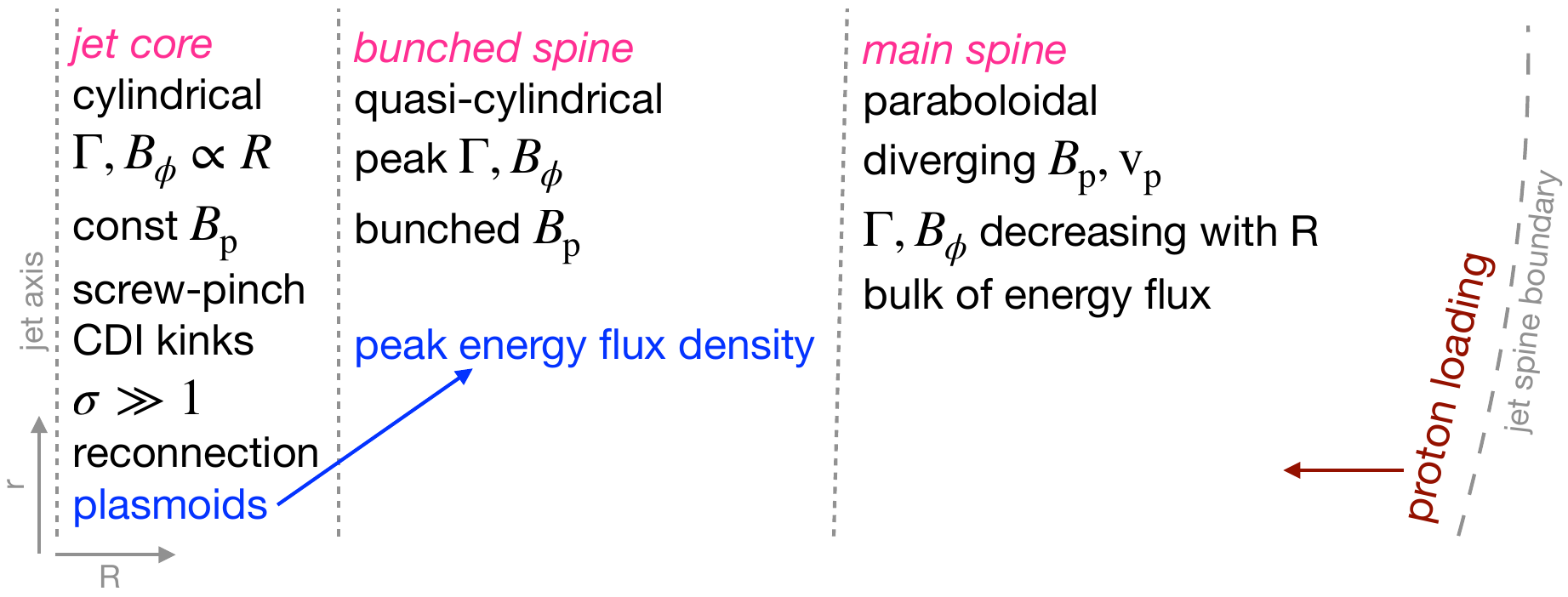}
\caption{The proposed lateral structure (not to scale) of a relativistic jet differentiated due to poloidal field bunching. The "bunched spine" zone is introduced as the region maximizing the jet energy flux density. Introduction of plasmoids from reconnection layers created due to current-driven instability in the jet core allows to multiply the energy density enhancement factors of the plasmoids and of the bunched jet spine.}
\label{fig_sketch_jet_structure}
\end{figure*}

The lateral structure of a relativistic jet spine in the differentiated ACZ stage is summarised schematically in Figure \ref{fig_sketch_jet_structure}.
Across such jet spine, we distinguish three zones: the innermost cylindrical {jet core}, the intermediate quasi-cylindrical {bunched spine}, and the outermost paraboloidal {main spine}.
The bunched spine is distinguished as the location of peak bulk Lorentz factor $\Gamma$ and toroidal magnetic field $B_\phi$, which means peak energy (Poynting) flux density.
The main spine zone is the most extended and carries the bulk of energy flux, it can be polluted by protons from the jet sheath via boundary instabilities \citep{2019MNRAS.490.2200C}, but this reduces the bulk Lorentz factor and dilutes the energy flux density.
The jet core is prone to the current-driven instability (CDI) \citep{1954RSPSA.223..348K,1982RvMP...54..801F,1998ApJ...493..291B}, which can accelerate particles via reconnection in high-$\sigma$ conditions \citep{2018PhRvL.121x5101A,2020ApJ...896L..31D,2022ApJ...931..137O}.
However, the reduced bulk Lorentz factor of the jet core would reduce the relativistic boost of associated radiation.

Efficient poloidal field bunching would result in a narrow bunched spine with radius of peak toroidal field strength $R_{\rm peak} \ll R_{\rm j}$, it also implies enhancement of peak Poynting flux density $S$ compared with the mean value across the jet spine.
In Appendix \ref{app_bunching}, we consider a generic model for radial profiles $S(R)$ and calculate the dependence between the enhancement factor $f_S = S_{\rm peak}/\left<S\right>$ and $R_{\rm peak}/R_{\rm j}$.
We find that $f_S \propto (R_{\rm peak}/R_{\rm j})^{-3/2}$, e.g., enhancement by factor $f_S \sim 10$ requires $R_{\rm peak}/R_{\rm j} \sim 0.15$.

In numerical simulations of the CDI, perturbations tend to spread from the most unstable toroidal field core to the outside \citep{2016MNRAS.456.1739B, 2019ApJ...884...39B, 2022ApJ...931..137O}.
Hence, we propose that plasmoids from reconnection layers induced by CDI in the jet core may spread to the bunched spine zone, as indicated in Figure \ref{fig_sketch_jet_structure}.
During that process, they would be need to be carried by the poloidal bulk flow of the jet in order to relativistically boost the emitted radiation.
Carrying large plasmoids by background high-$\sigma$ flow is not expected to be a highly efficient process, because of their significant inertia, which in a slightly different context results in anti-correlation between plasmoid growth and bulk acceleration \citep{2016MNRAS.462...48S,2020MNRAS.497.1365O}.
Moreover, the poloidal velocity shear is rarely accounted for in studies of CDI \citep{2012MNRAS.427.2480N}, and its effect is still unclear.
Alternatively, strongly sheared magnetic perturbations may induce reconnection layers directly in the bunched spine zone.

\section{Discussion and conclusions}
\label{sec_concl}

We considered an analogy between the lateral structures of relativistic jets and plasmoids produced by magnetic reconnection.
In both cases a key role is played by the toroidal component of magnetic field, the tension of which introduces an anisotropic stress that in principle can compress the energy density of both plasma and magnetic field.

Compression of energy density in reconnection plasmoids was investigated by kinetic particle-in-cell (PIC) numerical simulations in 2D and 3D periodic domains, starting from hot Harris layers and evolving to relaxed monster plasmoids.
We considered the effects of domain size, background magnetisation and guide magnetic field.
In our simulations without guide field, the plasma energy density achieves values $u_{\rm pl} \sim 10^2 u_{\rm B,0}$ (relative to the initial background magnetic energy density), higher than the magnetic energy density $u_{\rm B} \sim 10 u_{\rm B,0}$.
The highest energy densities correspond to the secondary plasmoids emerging between merging primary plasmoids.
However, the initial plasma energy density extends to $\sim 4 u_{\rm B,0}$ for the hot drifting particles.
In the presence of guide field or third dimension, the plasma energy densities are reduced by factors of a few.

We have also re-analysed the results of previous PIC simulations, first presented in \cite{2020MNRAS.497.1365O}, which used open boundaries and synchrotron cooling.
Those results show power-law tails in the distributions of both $u_{\rm pl}$ and $u_{\rm B}$ with indices close to $-5/3$.
The cooling efficiency affects mainly the extent of $u_{\rm B}$ distribution, which may explain the apparent lack of such power-law tail in the results of our new simulations.

In relativistic jets, the effect of toroidal magnetic field has been controversial, eventually it was understood that its tension is insufficient to collimate such jets.
However, toroidal field tension has a subtle effect of gradually differentiating the lateral jet structure.
The key effect of poloidal field bunching has been identified by \cite{2009ApJ...699.1789T}.
We propose that this effect forms an intermediate layer across the jet spine, referred to as the {bunched spine}, adjacent to the central {jet core}, corresponding roughly to the peaks of both toroidal field strength $B_\phi(R)$ and bulk Lorentz factor $\Gamma(R)$.
The bunched spine is characterised by enhanced Poynting flux density $S$.
For generic radial profiles of $S(R)$, the enhancement factor $f_S = S_{\rm peak}/\left<S\right>$ is sensitive to the relative peak radius $R_{\rm peak}/R_{\rm j}$.
Poloidal field bunching was demonstrated in 2D GRMHD global simulations of relativistic jets \citep{2019MNRAS.490.2200C}, however, it appears to be less efficient in 3D simulations.

Additional enhancement of luminosity of short flares of energetic radiation can be provided by the effect of kinetic beaming \citep{2012ApJ...754L..33C}, also known as the pitch-angle anisotropy \citep{2021PhRvL.127y5102C,2023ApJ...946L..51S}.
This effect is not expected to be important in relaxed reconnection plasmoids, but rather during the formation of plasmoid chains, and likely also during plasmoid mergers.
Our PIC simulations confirm that plasmoid mergers produce the highest energy densities in reconnection layers, and the resulting radiation intensity could be further enhanced by kinetic beaming.

The possibility to locally concentrate the momentum flux density of relativistic jets is important also for alternative models of rapid blazar flares, e.g., obliteration of stars wandering into the jet \citep[e.g.,][]{1997MNRAS.287L...9B}.
The efficiency of bulk acceleration can be increased in sharply inhomogeneous jet due to impulsive acceleration \citep{2011MNRAS.411.1323G}, which can also be described as a {relativistic whip}.

\subsection{The jet of M87}
The proposed scenario can be confronted with observations of relativistic AGN jets resolved by very-long-baseline interferometry (VLBI).
In the best-studied case of M87 \citep[for review see][]{2024A&ARv..32....5H}, the approaching jet (seen at the viewing angle of $\simeq 18^\circ$) has {long been} known to be limb-brightened at the scales\footnote{At the estimated distance of $D_{M87*} \simeq 16.8\,{\rm Mpc}$ and BH mass $M_{M87*} \simeq 6.5\times 10^9 M_\odot$ \citep{2019ApJ...875L...6E}, $1\,{\rm mas}$ corresponds to a projected linear scale of $\simeq 81\,{\rm mpc} \equiv 260 R_{\rm g}$, and a de-projected scale of $\simeq 0.26\,{\rm pc}$.} of $\sim 50\,{\rm mas}$ \citep{1989ApJ...336..112R}, $\sim 1\,{\rm mas}$ \citep{1999Natur.401..891J}, $\sim 0.2\,{\rm mas}$ \citep{2018A&A...616A.188K}.
The $<0.6\,{\rm mas}$ limbs were modeled by \cite{2022ApJ...936...79P,2023A&A...679L...1P,2024A&A...685L...3P} in terms of thick tubular jet sheath.
Recently, a third central ridge has been reported on the scales of $\sim 0.5\,{\rm mas}$ \citep{2023Natur.616..686L}, $\sim 8\,{\rm mas}$ \citep{2023Galax..11...39T}, possibly also at $\sim 300\,{\rm mas}$ \footnote{Presented by T. Savolainen at the IAU Symposium 375 in Kathmandu, Nepal (Dec 2022).}.
The $0.5\,{\rm mas}$ ridge was interpreted by \cite{2024ApJ...962...18L} in terms of strong beaming due to instant bulk acceleration of pair-dominated hollow-cone jet.
The central ridge appears more likely to be a signature of the jet core weakly beamed due to reduced bulk Lorentz factor.
Highly relativistic jet spine would be de-beamed at such large viewing angle.

According to the AGN unification paradigm \citep{1995PASP..107..803U}, an Fanaroff-Riley (FR) type I radio galaxy like M87 should be a misaligned counterpart of a BL Lac type blazar.
Beamed non-thermal blazar emission should be produced in a highly dissipative blazar zone.
Models of BL Lac type blazars do not provide strong constraints on the location.
A de-projected distance of $1\,{\rm pc}$ corresponds to $\simeq 4\,{\rm mas}$, indicating the $\sim 8\,{\rm mas}$ ridge as potentially adjacent to a blazar zone in the form of bunched spine.

A well known dissipative zone in the M87 jet is the HST-1 knot observed at $\simeq 860\,{\rm mas}$ \citep{2012A&A...538L..10G} with evidence for apparently superluminal motions \citep{1999ApJ...520..621B,2007ApJ...663L..65C} produced a spectacular X-ray/optical/radio outburst on the time scale of a few years \citep[e.g.,][]{2006ApJ...640..211H,2009ApJ...699..305H}.
It has been suggested to be the blazar zone of the M87 jet \citep{2014ApJ...788..139B}, which would pose severe problems of sufficient energy density and short variability timescale at deprojected distance of $\sim 220\,{\rm pc}$.

\subsection{Conclusions}
In the multi-scale system of a relativistic jet hosting plasmoids produced by magnetic reconnection, we consider it more likely to achieve a significant (by factor $\gtrsim 10$) energy density enhancement due to tension of toroidal magnetic field on the local scale of plasmoid cores rather than on the global scale of bunched jet spine (the layer adjacent to the central jet core).
Even a moderate ($\sim 3$) global-scale enhancement combined with a stronger ($\sim 10$) local-scale enhancement would boost the luminosity of rapid blazar flares by factor $\sim 30$ at standard jet bulk Lorentz factors of $\Gamma_{\rm j} \sim 10-20$, strongly reducing the extreme energetic requirements of such flares.

\begin{acknowledgement}
We gratefully acknowledge Poland's high-performance Infrastructure PLGrid (HPC Centers: ACK Cyfronet AGH, PCSS, CI TASK, WCSS) for providing computer facilities and support within computational grants no. PLG/2023/016444 and PLG/2024/017356;
and the Nicolaus Copernicus Astronomical Center for providing the Chuck cluster.
This work was supported by the Polish National Science Centre grant 2021/41/B/ST9/04306.
\end{acknowledgement}


\begin{appendix}

\section{Energetics of blazar flares}
\label{app_blazar_flares}

Rapid and luminous blazar flares require efficient conversion of very high energy densities.
For a typical estimate, consider a compact flaring region of radius $R_{\rm fl}'$ propagating with speed $\beta_{\rm fl} = v_{\rm fl}/c$ and Lorentz factor $\Gamma_{\rm fl} = (1-\beta_{\rm fl}^2)^{-1/2}$ (defining reference frame $\mathcal{O}'$), and observed at viewing angle $\theta_{\rm obs} \simeq 1/\Gamma_{\rm fl}$ corresponding to a Doppler factor $\mathcal{D}_{\rm fl} = [\Gamma_{\rm fl}(1-\beta_{\rm fl}\cos\theta_{\rm obs})]^{-1} \sim \Gamma_{\rm fl}$ resulting in a luminosity boost\footnote{Propagating pattern, rather than stationary \citep{1997ApJ...484..108S}.} $L_{\rm fl,obs} = \mathcal{D}_{\rm fl}^4 L_{\rm fl}'$ and shortening of variability time scale $t_{\rm fl} = t_{\rm fl}'/\mathcal{D}_{\rm fl}$.
In blazars, relativistic motions with bulk Lorentz factors $\Gamma_{\rm fl} \sim 20$ can be deduced from apparently superluminal proper motions with speeds $v_{\rm app} \sim 20c$ (with $c$ the speed of light) along the jet \citep[e.g.,][]{2001ApJS..134..181J,2016AJ....152...12L}.
The intrinsic parameters can be approximated as $t_{\rm fl}' \sim R_{\rm fl}'/c$ and $L_{\rm fl}' \sim 4\pi R_{\rm fl}'^2 u_{\rm fl}'c$ with $u_{\rm fl}'$ the intrinsic radiation energy density.
The total power carried by the flare photons can be estimated as $P_{\rm fl} \sim \pi R_{\rm fl}^2\Gamma_{\rm fl}^2u_{\rm fl}'c \sim [\Gamma_{\rm fl}^2/(4\mathcal{D}_{\rm fl}^4)] L_{\rm fl,obs} \sim L_{\rm fl,obs}/(4\Gamma_{\rm fl}^2)$ with $R_{\rm fl} = R_{\rm fl}' \sim \mathcal{D}_{\rm fl}ct_{\rm fl}$.
That power can be compared with the Eddington luminosity of the supermassive black hole (BH) $L_{\rm Edd} \simeq 1.5\times 10^{47}(M_{\rm BH}/10^9 M_\odot)\,{\rm erg/s}$ with $M_{\rm BH} \sim 10^9 M_\odot$ the BH mass in units of the solar mass $M_\odot$.
For flaring FSRQs this gives
$P_{\rm fl,FSRQ} \sim 0.4 (\Gamma_{\rm fl}/20)^{-2} L_{\rm Edd}$,
which is dangerously close to $L_{\rm Edd}$, and this is one reason to consider significantly higher values of $\Gamma_{\rm fl}$ \citep{2016ApJ...824L..20A}.
For flaring HBLs,
$P_{\rm fl,HBL} \sim 4\times 10^{-4} (\Gamma_{\rm fl}/20)^{-2} L_{\rm Edd}$,
more comfortable at least in terms of $L_{\rm Edd}$.

Consider that this compact blazar zone is only a small part of a relativistic jet with a total power $P_{\rm j}$ and Lorentz factor $\Gamma_{\rm j}$.
At distance scale $r$ the jet has lateral radius $R_{\rm j} = \theta_{\rm j} r$ with $\theta_{\rm j}$ the half-opening angle.
Interferometric radio observations of pc-scale blazar jets suggest that $\Gamma_{\rm j}\theta_{\rm j} \sim 0.2$ \citep{2013A&A...558A.144C}.
The jet can be subdivided laterally into streamtubes, of which we are interested only in the streamtube of radius $R_{\rm st} = R_{\rm fl}$ that completely and minimally contains the flaring region.
Neglecting the poloidal line bunching and assuming that power per unit solid-angle is roughly uniform across the jet, the fraction of jet power $P_{\rm st}$ in that streamtube would scale as $P_{\rm st}/P_{\rm j} \simeq (R_{\rm st}/R_{\rm j})^2$.
We can ask, at what distances is the jet power passing through a flaring region sufficient to power the observed flare, $P_{\rm j\to fl} > P_{\rm fl}$?
We obtain the following estimate:
\begin{equation}
r < ct_{\rm fl}\frac{2\mathcal{D}_{\rm fl}^3}{\Gamma_{\rm fl}} \frac{\Gamma_{\rm j}}{\Gamma_{\rm j}\theta_{\rm j}} \sqrt{\frac{P_{\rm j}}{L_{\rm fl,obs}}}\,.
\end{equation}
In the case of FSRQs we set $\Gamma_{\rm j} = 20$ and obtain $r < 0.02(\Gamma_{\rm fl}/20)^2 (P_{\rm j}/L_{\rm Edd})^{1/2}\,{\rm pc}$.
This is shorter from constraints using other methods, which typically find $r_{\rm FSRQ} \sim 0.1\;{\rm pc}$ \citep{2014ApJ...789..161N}.
This is another reason to increase the Lorentz factor of the flaring region.
Setting $\Gamma_{\rm fl} = 50$ \citep{2016ApJ...824L..20A} would result in $P_{\rm fl,FSRQ} \sim 0.07 L_{\rm Edd}$ and $r < 0.11\;{\rm pc}$.
This is particularly challenging for FSRQs producing $\sim 100\;{\rm GeV}$ flares like PKS 1222+216, for which pair-production absorption by optical/UV broad emission lines would be important at distances $r \lesssim 0.5\;{\rm pc}$ \citep{2012MNRAS.425.2519N}.
In the case of HBLs, $r < 0.17(\Gamma_{\rm fl}/20)^2 (P_{\rm j}/L_{\rm Edd})^{1/2}\,{\rm pc}$.

\section{Scale separation in relativistic jets}
\label{app_scale_separ}

Relativistic jets are characterised by very large separation between their macroscopic global scales (e.g., the lateral jet radius $R_{\rm j}$) and the microscopic plasma scales (e.g., the gyroradius $R_{\rm L}'$, formally in the jet co-moving frame).
A typical relativistic AGN jet with Lorentz factor $\Gamma_{\rm j} \sim 20$ and half-opening angle $\theta_{\rm j} \sim 0.2/\Gamma_{\rm j} \sim 0.01$ \citep{2013A&A...558A.144C} at the distance of $r \sim 0.1\;{\rm pc}$ (typical for blazar zones; \citealt{2014ApJ...789..161N}) would have a radius of $R_{\rm j} \sim 10^{-3}\,{\rm pc} \simeq 3\times 10^{15}\,{\rm cm}$ and may contain magnetic fields of co-moving strength $B' \sim 2.7\;{\rm G}$\footnote{Corresponding to the magnetic jet power of $P_{\rm B} = (c/8)(\Gamma_{\rm j}R_{\rm j}B')^2 \sim 10^{44}\,{\rm erg\,s^{-1}}$.}, and electrons of mean random Lorentz factor $\gamma_{\rm e} \sim 100$ having a gyroradius of $R_{\rm L}' = \gamma_{\rm e}m_{\rm e}c^2/eB' \sim 6\times 10^4\,{\rm cm}$, resulting in a huge scale separation of $R_{\rm j}/R_{\rm L}' \sim 5\times 10^{10}$.
This would mean an enormous potential for energy density enhancement, if it could be fully utilised.

The presence of compact jet core would introduce an intermediate scale -- the core radius $R_{\rm c} \sim 6M_{\rm BH}$ (for high BH spin $a \simeq 0.9$).
For a BH mass of $M_{\rm BH} \sim 10^9 M_\odot \sim 1.5\times 10^{14}\,{\rm cm}$, one could expect $R_{\rm c} \sim 10^{15}\,{\rm cm}$ (and $R_{\rm j} \sim 20 M_{\rm BH}$ at $r \sim 2000 M_{\rm BH}$), implying $R_{\rm j}/R_{\rm c} \sim 3$.
This ratio could be increased by considering a lower $M_{\rm BH}$ or a broader, less relativistic jet, but it can hardly be much larger than $\sim 10$, and thus large energy density enhancement should not be expected to result from a compact jet core.
Global GRMHD numerical simulations of jets reaching distances of $\sim 10^3 M_{\rm BH}$ with well resolved jet spine of radius $R_{\rm sp}$, both in 2D \citep{2019MNRAS.490.2200C} and in 3D \citep{2022ApJ...924L..32R,2024MNRAS.533..254S}, show bulk Lorentz factors $\Gamma_{\rm j}$ monotonically increasing across the jet from its axis until at least $\sim 0.4 R_{\rm sp}$, which is consistent with toroidal magnetic field peaking at $\gtrsim R_{\rm sp}/2$, which implies no significant energy density enhancement due to tension of global toroidal field.

In magnetic reconnection involving macroscopically large ($\sim R_{\rm j}$) and microscopically thin ($\sim R_{\rm L}'$) current layers, plasmoids will be produced and grown to a broad range of radii $R_{\rm pl}$, potentially spanning the entire available range of scales $R_{\rm L}' < R_{\rm pl} < R_{\rm j}$.
For plasmoids produced during relativistic magnetic reconnection, our PIC simulations with initial scale separation up to $L/\rho_0 = 3600$ produced relaxed monster plasmoids with scale separation of $R_{\rm out}/R_{\rm c} \simeq 13$ and with contrast of total energy density of $u_{\rm tot}(R_{\rm c}) / u_{\rm tot}(R_{\rm out}) \simeq 16$, much less than the idealised expectation of $(R_{\rm out}/R_{\rm c})^2 \simeq 170$.
However, in a hypothetical case of large-scale relativistic reconnection producing a plasmoid reaching a size approaching the jet radius, e.g., $R_{\rm out} \simeq L/2 \sim R_{\rm j}/10$ and initial scale separation of $L/R_{\rm L}' \sim 10^{10}$, assuming that the plasmoid core radius scales like $R_{\rm c}/L \sim 5.6\times 10^{-4}(L/R_{\rm L}')^{1/2}$, this would imply a scale separation of $R_{\rm out}/R_{\rm c} \sim 8\times 10^7$, and idealised energy density enhancement up to $\sim 6\times 10^{15}$.
Such large plasmoids were considered by \cite{2019MNRAS.486.1548M} to explain rapid gamma-ray flares of blazars in terms of synchrotron self-Compton emission, assuming uniform energy density, which in the context of this work appears rather unfortunate.

\section{Model for bunching momentum flux across jet}
\label{app_bunching}

As discussed in Section \ref{sec_jets}, bunching (internal collimation) of poloidal magnetic field towards the jet axis \citep{2009ApJ...699.1789T} is expected to bunch the Poynting (electromagnetic momentum) flux into the bunched spine layer.
Shifting inwards the radius $R_{\rm peak}$ of peak momentum flux density relative to the jet (spine) radius $R_{\rm j}$ affects the mean momentum flux density across the jet.
By considering a radial profile $S(R)$ of momentum flux density, one can calculate the ratio of peak value $S_{\rm peak} = S(R_{\rm peak})$ to the mean value $\left<S\right> = (2/R_{\rm j}^2)\int_0^{R_{\rm j}}R S(R)\,{\rm d}R$ as the {jet momentum enhancement factor} $f_S = S_{\rm peak}/\left<S\right>$.
In the highly magnetised jet core ($R < R_{\rm peak}$), the momentum flux is dominated by the Poynting flux of density $S(R) \simeq (c/4\pi)v_z(R) B_\phi^2(R)$.
The toroidal field should satisfy $B_\phi(R \ll R_{\rm j}) \propto R$ in order to assure a finite poloidal current density $j_z(R) = (c/4\pi R)\,{\rm d}(RB_\phi)/{\rm d}R$ and a uniform magnetic pitch $\mathcal{P}(R) = R B_z(R) / B_\phi(R)$ for $B_z(R \ll R_{\rm j}) \simeq {\rm const}$.
Even in the case of uniformly accelerated core with $v_z(R \ll R_{\rm j}) \simeq {\rm const}$, the innermost momentum flux density should scale like $S(R \ll R_{\rm j}) \propto R^2$.

\begin{figure}
\includegraphics[width=\columnwidth]{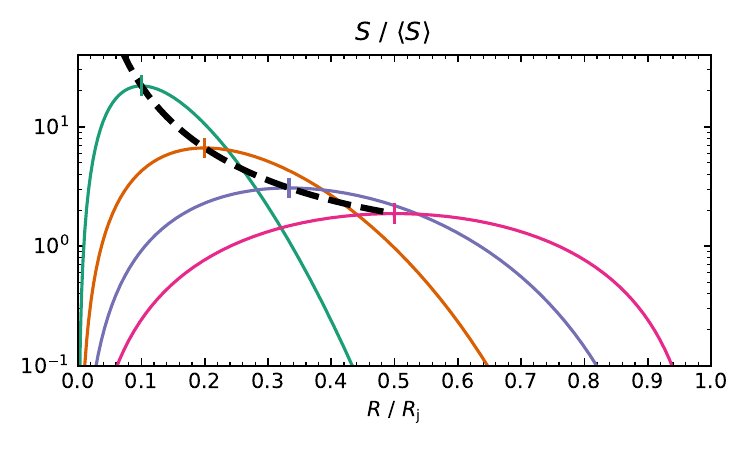}
\caption{Color solid lines show examples of model radial profiles across a cylindrical jet of radius $R_{\rm j}$ of momentum flux density $S(R)$ (defined by Eq. \ref{eq_power_rprof}) normalised by the mean value $\left<S\right>$ (Eq. \ref{eq_mean_power}) for $p = 2$ and different values of $q$ (corresponding to $R_{\rm peak}/R_{\rm j} = 1/10, 1/5, 1/3, 1/2$). The thick dashed line shows the enhancement factor $f_S = S_{\rm peak}/\left<S\right>$ (Eq. \ref{eq_enhancement}) as function of $R_{\rm peak}/R_{\rm j}$.}
\label{fig_power_rprofs}
\end{figure}

Consider a mathematical model of momentum flux density profile:
\begin{equation}
S(R) = S_0\, (R/R_{\rm j})^p\, (1-R/R_{\rm j})^q
\label{eq_power_rprof}
\end{equation}
The peak radius is given by $R_{\rm peak}/R_{\rm j} = p/(p+q)$, corresponding to $S_{\rm peak} = S_0 p^p q^q / (p+q)^{p+q}$.
We choose $p = 2$ to satisfy the inner scaling, and $q \ge 2$ in order to have $R_{\rm peak} \le R_{\rm j}/2$.
Substituting $r = R/R_{\rm j}$, one can integrate the mean value:
\begin{equation}
\frac{\left<S\right>}{2S_0} = \int_0^1r^{p+1}(1-r)^q\,{\rm d}r
= \frac{\Gamma(p+2)\Gamma(q+1)}{\Gamma(p+q+3)}
\label{eq_mean_power}
\end{equation}
The peak enhancement factor is thus:
\begin{equation}
f_{\rm S} = \frac{S_{\rm peak}}{\left<S\right>}
= \frac{p^p q^q\,\Gamma(p+q+3)} {2(p+q)^{p+q}\,\Gamma(p+2)\Gamma(q+1)}\,.
\label{eq_enhancement}
\end{equation}
Examples of profiles $S(R)/\left<S\right>$ and the function $f_{\rm S}(R_{\rm peak})$ for $p = 2$ are presented in Figure \ref{fig_power_rprofs}.
The case without poloidal field bunching ($q = 2$) corresponds to $R_{\rm peak} = R_{\rm j}/2$ and is characterised by small enhancement of $f_S \simeq 1.9$.
However, as $R_{\rm peak}/R_{\rm j}$ decreases due to stronger bunching, $f_{\rm S}$ increases, e.g., to $\simeq 6.6$ for $R_{\rm peak} = R_{\rm j}/5$, to $\simeq 22$ for $R_{\rm peak} = R_{\rm j}/10$.
The enhancement function for $p=2$ and $0.1 < R_{\rm peak}/R_{\rm j} < 0.5$ can be well approximated as $f_{\rm S} \simeq 0.66(R_{\rm peak}/R_{\rm j})^{-3/2}$.

As will be shown in a different article, in the case without bunching ($q = 2$) such profile makes a good approximation to the radial distribution of poloidal Poynting flux in relativistic jets obtained in "extreme-resolution" GRMHD simulation of \citep{2022ApJ...924L..32R} reaching distances of $\sim 10^3 R_{\rm g}$ in 3D.
The cases with bunching need to be compared with GRMHD simulations reaching larger distances, e.g., $\sim 10^5 R_{\rm g}$ in 2D \citep{2019MNRAS.490.2200C}.

\end{appendix}

\end{document}